%% file: main.tex
\documentclass[12pt]{article}

\usepackage{amsmath,amsfonts,amssymb}
\usepackage{amscd}
\usepackage{}
\usepackage{color}
\usepackage{comment}
\usepackage{cite}
\usepackage[utf8]{inputenc}

\newlength{\minitwocolumn}
\newcommand\bcQ{\mbox{\boldmath $Q$}}

\newcommand\undert{\underline{t}}
\newcommand\underalpha{{\underline{\alpha}}}
\newcommand\unders{{\underline{s}}}
\newcommand\undershm{{\underline{s}{}_{HM}}}
\newcommand\undern{{\underline{n}}}

\newcommand\tila{{\tilde{a}}}
\newcommand\tiltila{{\tilde{a}_{(-1)}}}

\newcommand{\he}{\hat{\epsilon}}
\newcommand{\hm}{\hat{\mu}}
\newcommand{\calP}{{\cal P}}
\newcommand{\dha}{\hat{\delta}}

\newcommand\bcA{\mbox{\boldmath $A$}}
\newcommand\bcB{\mbox{\boldmath $B$}}
\newcommand\bcC{\mbox{\boldmath $C$}}
\newcommand\bcD{\mbox{\boldmath $D$}}
\newcommand\bcd{\mbox{\boldmath $d$}}

\newcommand\bcZ{\mbox{\boldmath $Z$}}

\newcommand\bcF{\mbox{\boldmath $F$}}
\newcommand\bcH{\mbox{\boldmath $H$}}

\newcommand\tildes{{\tilde{s}}}
\newcommand\tilden{{\tilde{n}}}

\newcommand\gvee{{w}}
\newcommand\hvee{{v}}

\def\bC{{\mathbb{C}}}

\newcommand{\bracket}[2]{\langle #1,\,#2\rangle}

\def\bc#1{\textcolor{blue}{#1}}

\begin{document}
\begin{titlepage}
\begin{flushright}
\null \hfill Preprint TU-1018\\[3em]
\end{flushright}

\begin{center}
{\Large \bf
Higher Gauge Theories from Lie n-algebras 
and Off-Shell Covariantization
}
\vskip 1.2cm

Ursula Carow-Watamura${}^{a,}$\footnote{E-mail:\
ursula@tuhep.phys.tohoku.ac.jp}, Marc Andre Heller${}^{a,}$\footnote{E-mail:\
heller@tuhep.phys.tohoku.ac.jp}, Noriaki Ikeda${}^{b,}$\footnote{E-mail:\
nikeda@se.ritsumei.ac.jp}, Yukio Kaneko${}^{a,}$\footnote{E-mail:\
y\textunderscore kaneko@tuhep.phys.tohoku.ac.jp} 
%
~and Satoshi Watamura${}^{a,}$\footnote{
E-mail:\ watamura@tuhep.phys.tohoku.ac.jp}

\vskip 0.4cm
{

\it
${}^a$
Particle Theory and Cosmology Group, \\
Department of Physics, Graduate School of Science, \\
Tohoku University \\
Aoba-ku, Sendai 980-8578, Japan \\ 

\vskip 0.4cm
${}^b$
Department of Mathematical Sciences,
Ritsumeikan University \\
Kusatsu, Shiga 525-8577, Japan \\

}
\vskip 0.4cm



\begin{abstract}
We analyze higher gauge theories in various dimensions using a supergeometric method based on 
a differential graded symplectic manifold, called a QP-manifold, which is closely related to the BRST-BV formalism in gauge theories.
Extensions of the Lie 2-algebra gauge structure are formulated within the Lie n-algebra induced by the QP-structure. 
We find that in $5$ and $6$ dimensions there are special  
extensions of the gauge algebra. In these cases, a restriction of 
 the gauge symmetry by imposing constraints on the
auxiliary gauge fields leads to a covariantized theory.  As an example we show that we can obtain an off-shell covariantized higher gauge theory in $5$ dimensions, which is similar to the one proposed in \cite{Ho:2012nt}.

\end{abstract}
\end{center}
\end{titlepage}

\newpage

\setcounter{tocdepth}{2}
\tableofcontents

\input intro.tex

\input sec02.tex

\input sec03.tex

\input sec04.tex

\input sec04-2.tex

\input sec05.tex
\input discussion.tex

\appendix

\input app01.tex

\input app02.tex


\bibliographystyle{h-physrev}
\bibliography{reference}
\end{document}

%% file: intro.tex
\end{comment}

\section{Introduction}
The dynamical objects of M-theory are M2- and M5-branes. Two M5-branes interact
by M2-branes extending between them. Their intersections are the one-dimensional boundaries of the M2-branes, which form self-dual strings charged under the self-dual B-field. These are soliton solutions of the so-called 6-dimensional ${\cal N}=(2,0)$ theory. This theory is believed to encode parallel transport of self-dual strings and therefore to be related to gauge theories containing higher form gauge fields.
Such higher gauge theories are considered as candidates to describe the effective dynamics of multiple M5-brane systems and discussed from various perspectives: 

The worldvolume effective theory of multiple M5-branes is described by the six-dimensional ${\cal N}=(2,0)$ supersymmetric theory \cite{Witten:1995zh,Witten:2009at}.
The authors of \cite{Lambert:2010wm} constructed a nonabelian on-shell ${\cal N}=(2,0)$ tensor multiplet, which is based on a $3$-algebra gauge structure. 
This direction was further investigated in \cite{Bonetti:2012st}, where a 5-dimensional superconformal action was proposed, as a candidate to study the dynamics of ${\cal N}=(2,0)$ self-dual nonabelian tensors in 6 dimensions. 
In \cite{Chu:2012um} an action for a nonabelian 2-form in 6 dimensions, whose equation of motion gives a self-duality constraint on the field strength, was proposed. It contains manifest Lorentz symmetry in 5 dimensions and upon dimensional reduction it leads to 5-dimensional Yang-Mills theory accompanied with some higher derivative corrections.
As for supersymmetrization of higher gauge theories, the authors of \cite{Ho:2014eoa} proposed a supersymmetric nonabelian self-dual gauge theory of 2-form fields in 6 dimensions, which reduces to a single M5-brane, if the gauge group is abelian.
\cite{Ritter:2015zur} considers a generalization of higher gauge theory, which models finite gauge transformations encoded in principal 2-bundles on 2-spaces (categorified spaces). The authors argue, that the 3-Lie algebra model of the 6-dimensional ${\cal N}=(2,0)$ worked out in \cite{Lambert:2010wm} can be interpreted in their proposed generalized higher gauge theory setting.
In \cite{Ho:2011ni} the authors worked out a closed and nonabelian gauge algebra for a chiral 2-form potential in 6 dimensions, with one spatial direction compactified on a circle. It was shown that the resulting transformation law is nonlocal along the circle direction, but reduces to Yang-Mills theory in 5 dimensions for small circle radia.

From these investigations we understand that there are two main obstructions to obtain a local action with manifest Lorentz symmetry \cite{Chu:2012um,Ho:2011ni,Ho:2014eoa}. One originates from the difficulty to formulate the theory with nontrivially interacting tensor field and the other is the difficulty to construct the action for the self-dual tensor field \cite{Perry:1996mk,Pasti:1996vs, Henneaux:1997ha, Bekaert:2000qx}. In this paper, 
we focus on the first problem.

Along this line, the authors of \cite{Baez:2010ya} have proposed a generalization of parallel transport of point-like objects to parallel transport of string-like objects.
This higher parallel transport leads to a gauge theory of a 2-form gauge field.
In their approach, the surface swept out by one-dimensional objects is reparametrization invariant, allowing for the introduction of a so-called 
Wilson surface, 
the higher analogue of the Wilson line in Yang-Mills theory.
The requirement of the consistency of 
this Wilson surface 
directly leads to the so-called crossed module as governing structure.
This idea was further investigated in the context of semistrict Lie $2$-algebra structures \cite{Baez:2003fs}, which is a generalization of the differential crossed module.
Related self-dual string solutions were constructed in \cite{Palmer:2014jma}.

However, in the construction based on the differential crossed module, the 3-form field strength $H$ is not covariant under gauge transformations, unless the 2-form field strength is zero (called fake curvature condition). 
Then, as is discussed in \cite{Baez:2010ya}, an action with nonabelian gauge symmetry cannot be formulated except for a topological action of BF type.
As a result, the theory becomes topological or essentially free.

In \cite{Ho:2012nt} a modification of the gauge transformation law of a differential crossed module was proposed in order to circumvent the fake curvature condition. The higher field strengths defined in \cite{Ho:2012nt} transform covariantly under the modified gauge transformations without fake curvature condition, leading to topologically nontrivial nonabelian gerbes.

Our aim is to obtain a nonabelian, gauge symmetric, interacting and local field theory of a higher gauge field.
From the point of view of the action principle,
it is desirable to have a gauge invariant local action with a term quadratic in the field strengths,
to obtain a unitary theory after quantization.
However, so far a satisfactory theory has not been constructed.
One way to obtain a dynamical theory of 
a 2-form gauge field is to require off-shell covariance of 
the 3-form field strength under gauge symmetry,
i.e.,~covariance without fake curvature condition.
If we can define a gauge symmetry, that closes off-shell, with or without introducing auxiliary fields, we may write an interacting action.
Instead of starting from the action, one can analyze field strengths and equations of motion directly and require consistency under gauge symmetries
\cite{vanNieuwenhuizen:1982zf, D'Auria:1982pm}.
See also \cite{Bojowald:2004wu, Lavau:2014iva, Gruetzmann:2014ica}.
\subsection{Off-shell covariantization}
If the field strengths transform off-shell covariantly, i.e., by adjoint transformation without employment of the equations of motion,
the quadratic action is invariant under the gauge transformation.
In this paper, we analyze a way to construct such an off-shell covariant field strength, called 
\emph{off-shell covariantization} hereafter.

For this construction,
we use the supermanifold method on a so-called QP-manifold \cite{Schwarz:1992nx,Schwarz:1992gs}, which we explain in detail in section 2. 
We consider a QP-manifold ${\cal M}_n = T^{*}[n]{\cal N} = T^*[n](W[1] \oplus V[2])$, where $W$ and $V$ are vector spaces. In general, this structure induces a symplectic Lie n-algebra.
The QP-structure on this space includes the differential crossed module and the semistrict Lie 2-algebra.
Gauge fields, gauge transformations and field strengths are constructed by associating supercoordinates on the QP-manifold to fields on the spacetime $\Sigma$ \cite{Castellani,Fiorenza:2010mh,Lavau:2014iva}. 
Consistent field strengths and gauge symmetries are determined by a geometric datum of the corresponding QP-manifold, which is called Hamiltonian function
(also called homological function). However, in general,
gauge symmetries of field strengths are \emph{on-shell} covariant
and we cannot avoid the fake curvature condition.

The procedure of off-shell covariantization can be performed by the following steps.

\begin{enumerate}
\item By solving the master equation of the QP-manifold ${\cal M}_n$, 
we obtain relations among the structure constants, 
which induce a symplectic Lie n-algebra.
\item Derive the field strengths and gauge transformations according to
the standard procedure.
\item Covariantize by imposing an appropriate constraint on 
the conjugate auxiliary fields, which reduces the symplectic Lie n-algebra to a nontrivial extension of a Lie 2-algebra. We allow to impose additional constraints on
the structure constants, if necessary.
\item Investigate the remaining gauge symmetry.
Note that on-shell closure of the gauge algebra is guaranteed by construction.
\end{enumerate}
By taking a proper constraint, the reduced field strengths
become off-shell covariant under the residual gauge symmetry.

The organization of this paper is as follows. In section 2, 
we briefly introduce the concept of QP-manifolds used in this paper.
In section 3, 
we discuss an $(n+1)$-dimensional higher gauge theory 
based on a general QP-manifold structure on ${\cal M}_n$.
We consider canonical transformations 
on ${\cal M}_n$ to 
classify equivalent higher gauge theories for generic $n$. 
We see that for generic $n$, the theory is equivalent to a higher gauge theory
induced by a semistrict Lie 2-algebra.
We also find that we can consider extra terms in the Hamiltonian function 
of a QP-structure for $n \leq 5$.
In section 4, we discuss the $n =4$ case in detail.
We compare our results with the theory given 
in \cite{Ho:2012nt}.
This case is of particular interest for physics,
since it can be related to the multiple M5-brane system compactified on $S^1$ 
\cite{Ho:2012nt,Ho:2011ni,Ho:2014eoa}.
In section 5, we discuss the $n =5$ case.
Section 6 is devoted to discussion.

%% file: sec02.tex
\end{comment}

\section{
Higher gauge theory
and QP-manifolds
}
In this section, we briefly introduce higher gauge theories and
the QP-manifold method.
Then,
we examine Hamiltonian functions and possibilities of deformations of these theories. Finally, we explain canonical transformations of the QP-manifold as a preparation to discuss higher gauge structures.

\subsection{Higher gauge theory}

Higher gauge theories are characterized by the appearance of higher form gauge fields and their nontrivial interaction.
For our purpose in this paper, the existence of a 1-form gauge field $A$ and 2-form gauge field $B$ is sufficient. 

In ref.\cite{Baez:2010ya} (and references therein), the authors discussed the crossed module 
given by a pair of Lie groups $G, H$
corresponding to the gauge fields $A^a$ and $B^A$.
In addition to the operations on the two Lie groups, 
the crossed module contains two maps $t: H \rightarrow G$ 
and $\alpha: G \rightarrow \mathrm{Aut}(H)$,
which satisfy compatibility conditions. Here, we briefly review their construction.

Let $\mathfrak{g}$ and $\mathfrak{h}$
be the Lie algebras corresponding to $G$ and $H$.
The corresponding infinitesimal object is called 
a \emph{differential crossed module}, which is a pair of two Lie
algebras $\mathfrak{g}$ and $\mathfrak{h}$
with two homomorphisms 
$\undert: \mathfrak{h} \rightarrow \mathfrak{g}$
and $\underalpha: \mathfrak{g} \rightarrow \mathrm{Der}(\mathfrak{h})$,
corresponding to $t$ and $\alpha$.

Let $g_a \in \mathfrak{g}$ and
$h_A \in \mathfrak{h}$ be the bases of the respective Lie algebras with 
Lie brackets given by
\begin{align}
[g_{a},g_{b}]=&-f^{c}_{ab}g_{c},\
[h_{A},h_{B}]=\tilde{f}^{C}_{AB}h_{C},
\end{align}
where 
$f^{c}_{ab}$ and $\tilde{f}^{C}_{AB}$ are structure constants.
The maps $\undert$ and $\underalpha$ are defined as
\begin{align}
\undert(h_{A})&=t^{a}_{A}g_{a},
\\
\underalpha(g_{a})h_{A}&= \alpha_{aA}^{B}h_{B},
\end{align}
with coefficients $t^{a}_{A}$ and $\alpha_{aA}^{B}$.
For the relations between the structure constants we refer to appendix A.

A systematic derivation leads to the field strengths associated to the ordinary and the higher gauge field,
\begin{align}
	F^a &= dA^a - \frac{1}{2}f^a_{bc} A^b \wedge A^c - t^a_A B^A, \\
	H^A &= dB^A + \alpha_{aB}^A A^a \wedge B^B,
\end{align}
and their gauge transformations,
\begin{align}
	\delta A^a &= d\epsilon^a - f^a_{bc}A^b\epsilon^c + t^a_A \mu^A, 
\label{gaugetransfcrossedmodule1} \\
	\delta B^A &= d\mu^A + \alpha_{aB}^A A^a\wedge \mu^B - \alpha_{aB}^A\epsilon^a B^B,
\label{gaugetransfcrossedmodule2}
\end{align}
where $\epsilon^a$ and $\mu^A$ are ordinary and higher gauge parameter, respectively.

In general, one finds that the 3-form field strength is not covariant under gauge transformations, leading to the so-called fake curvature condition. Therefore, one cannot introduce the corresponding kinetic term in the Lagrangian of such higher gauge theories. This is the problem we want to address in this paper. We show how to get to off-shell covariantized higher gauge theory by an extension of the crossed module ansatz.

\subsection{QP-manifolds and canonical transformations}
A QP-manifold of degree $n$ is defined by a triple $({\cal M}, \omega, \Theta)$.
${\cal M}$ is an N-manifold, which is a graded manifold with 
a nonnegative $\mathbb{Z}$-grading.
$\omega$ is a graded symplectic form of degree $n$,
which induces a graded Poisson bracket $\{-,-\}$ of degree $-n$.
$\Theta$ is a function of degree $n+1$ on ${\cal M}$, which satisfies
the classical master equation, $\{\Theta, \Theta \}=0$,
\cite{Schwarz:1992nx,Schwarz:1992gs}.
$\Theta$ is called Hamiltonian function or homological function.
From this Hamiltonian, the homological vector field $\bcQ$ is 
obtained by
\begin{align}
\bcQ f=\{\Theta,f\},
\end{align}
where $f\in C^{\infty}(\mathcal{M})$.
The classical master equation is equivalent to the nilpotency condition on the homological vector field, ${\bcQ^2=0}$.



We consider graded manifolds ${\cal M}_{n} = T^*[n]{\cal N}=T^{*}[n](W[1]\oplus V[2])$ for $n\in\mathbb{N}$,
where $W$ and $V$ are vector spaces.\footnote{$W$ corresponds to $\mathfrak{g}^*$ and $V$ to $\mathfrak{h}^*$, respectively.} As we shall see, these vector spaces are related to the two vector spaces in the definition of the crossed module.

%
Let $q^{a}$ and $Q^{A}$ be local coordinates on $W[1]$ and $V[2]$ with degree $1$ and $2$, respectively. 
The degree is identified with the ghost number in the BRST-BV formalism of the corresponding field theory.
The conjugate coodinates with respect to the fiber $T^*[n]$
are denoted by $(p_{a},P_{A})$
and are of degree $(n-1, n-2)$.
Therefore, the local coordinates on ${\cal M}_{n}$ are
$(q^a, Q^A, p_{a}, P_{A})$ with degree $(1,2,n-1,n-2)$.
Coordinates of odd degree are Grassmann odd quantities. 

We consider 
the graded symplectic form $\omega$,
\begin{align}
\omega = (-1)^n \delta q^a \wedge \delta p_a
+ \delta Q^A \wedge \delta P_A.
\end{align}
The corresponding Poisson bracket $\{-,-\}$ on functions 
$f, g \in C^{\infty}({\cal M}_n)$
is given by 
\begin{equation}
\{f,g\}
= \frac{f\overleftarrow{\partial}}{\partial q^{a}}\frac{\partial g}{\partial p_{a}}+(-1)^{n}\frac{f\overleftarrow{\partial}}{\partial p_{a}}\frac{\partial g}{\partial q^{a}}+\frac{f\overleftarrow{\partial}}{\partial Q^{A}}\frac{\partial g}{\partial P_{A}}-\frac{f\overleftarrow{\partial}}{\partial P_{A}}\frac{\partial g}{\partial Q^{A}}.
\label{gradedPoissonbracket}
\end{equation}
Note that we define the right derivative by $\frac{f\overleftarrow{\partial}}{\partial X}=(-1)^{|X|(|f|-|X|)}\frac{\partial f}{\partial X}$,
where $|f|$ is the degree of the function $f$.

A canonical transformation $\delta_{\alpha}$ is defined by 
adjoint action of a function $\alpha$ of degree $n$ as
\begin{align}
e^{\delta_{\alpha}}f=f+\{f,\alpha\}+\frac{1}{2}\{\{f,\alpha\},\alpha\}+\cdots.\label{eq:defcanonical}
\end{align}
It preserves the Poisson bracket,
\begin{align}
\{e^{\delta_{\alpha}}f,e^{\delta_{\alpha}}g\}=e^{\delta_{\alpha}}\{f,g\},
\end{align}
for any function $f,g \in C^{\infty}({\cal M}_n)$. 
Since the generator of the canonical transformation $\alpha$ is of degree $n$, it is degree-preserving.

For details on the conventions related to QP-manifolds and graded differential calculus, 
we refer to
\cite{Bessho:2015tkk}.

\subsection{Higher gauge theory from QP-manifolds}\label{HGTfromQP}
We construct a gauge field theory
on an $(n+1)$-dimensional spacetime $\Sigma$
using the BV-AKSZ formalism \cite{Alexandrov:1995kv, Ikeda:2012pv}.
For this, we consider the graded manifold $T[1]{\Sigma}$ with local coordinates 
$(\sigma^{\mu},\theta^{\mu})$ of degree $(0,1)$,
where $\sigma^{\mu}$ are coordinates on the base manifold $\Sigma$ 
corresponding to the spacetime,
and $\theta^{\mu}$ are coordinates on the fiber.

The gauge fields are obtained by a pullback map
$a^*$ and a degree preserving map 
${\tila}^*$ as in 
\cite{Gruetzmann:2014ica}.
Given a map between graded manifolds,
$a: T[1]\Sigma \rightarrow {\cal M}_n$, 
the pullback of elements of $C^{\infty}({\cal M}_n)$
by $a^*$ gives superfields.
For example, a coordinate $z$ of degree $k$ on ${\cal M}_n$ induces a superfield of degree $k$,
\begin{align}
\bcZ(\sigma, \theta)& \equiv a^{*}(z) = \sum_{j=0}^{n+1} \frac{1}{j!} \theta^{\mu_1} \cdots \theta^{\mu_j} Z^{(j)}_{\mu_1 \cdots \mu_j}(\sigma).
\end{align}
We denote the degree by $|\bcZ| = k$.
We also denote the $j$-th component by
$\bcZ^{(j)}(\sigma, \theta) \equiv \frac{1}{j!} \theta^{\mu_1} \cdots \theta^{\mu_j} Z^{(j)}_{\mu_1 \cdots \mu_j}(\sigma)$.
This map automatically introduces corresponding gauge fields, ghosts and 
antifields in the BV formalism.

For the correspondence to physical fields, another degree, called \emph{form degree}, $\mathrm{deg} (\Phi)$ is introduced. We assign form degrees to $(\sigma^{\mu}, \theta^{\mu})$ as $(0,1)$.
$(|\Phi| - \mathrm{deg} (\Phi))$ is called \emph{ghost number}.

Since $\theta^{\mu}$ is of form degree $1$, $Z^{(j)}_{\mu_1 \cdots \mu_j}$ has ghost number $(k-j)$.
The ghost number zero component $Z^{(k)}_{\mu_1 \cdots \mu_k}$ is
a $k$-form gauge field. A positive ghost number component is a ghost and
a negative ghost number component is an antifield.
Especially, the ghost number $1$ part $Z^{(k-1)}_{\mu_1 \cdots \mu_{k-1}}$ 
is the gauge parameter for the field $Z^{(k)}_{\mu_1 \cdots \mu_k}$.

The super field strength corresponding to a coordinate $z$ is defined by 
\begin{align}
\bcF_Z \equiv \bcd \circ a^{*}(z) - a^{*} \circ \bcQ(z),
\label{superfeildstrength}
\end{align}
where $\bcd = \theta^{\mu} \partial_{\mu}$ is the superderivative.
The corresponding physical field strength
is the degree $|z|+1$ part of the super field strength
$\bcF_Z$,
\begin{align}
{\cal F}_z= (\bcd \circ a^{*}(z) - a^{*} \circ \bcQ(z)) |_{|z|+1},
\label{generalfieldstrength}
\end{align}
where $|_{|z|+1}$ denotes taking the degree $|z|+1$ part.

We can define a \emph{degree-preserving map},
$\tila: T[1]\Sigma \rightarrow {\cal M}_n$, such that,
for a function of degree $k$ on the target space, 
the pullback $\tila^*$ chooses the component of $k$-th order in $\theta^\mu$ in the superfield expansion, i.e.,
\begin{align}
\tila^{*}(z)& = \frac{1}{k!} d\sigma^{\mu_1} \wedge \cdots \wedge d\sigma^{\mu_k} 
Z^{(k)}_{\mu_1 \cdots \mu_k}(\sigma),
\label{expansionofsuperfield}
\end{align}
where we identify the degree $1$ coordinate $\theta^{\mu}$ with the basis of the differential forms $d\sigma^{\mu}$. 

A degree $k$ coordinate corresponds to a $k$-form gauge field.
For example,
\begin{align}
\tila^{*}(q^{a})&\equiv A^{a}=A^{a}_{\mu}d\sigma^{\mu},\\
\tila^{*}(Q^{A})&\equiv B^{A}=\frac{1}{2}B^{A}_{\mu\nu}d\sigma^{\mu}\wedge d\sigma^{\nu}.
\end{align}
The corresponding field strengths $F_z$
are defined by the map $F$:
\begin{align}
F_z\equiv F(z) 
= d \circ \tila^{*}(z) - \tila^{*} \circ \bcQ(z),
\label{generalfieldstrength}
\end{align}
i.e.,
\begin{align}
F^{a} &\equiv (d\tila^{*}-\tila^{*}\bcQ)q^{a}
=d \tila^{*}(q^{a})-\tila^{*}(\{\Theta,q^{a}\}),
\label{fieldstrengthF}\\
H^{A}&\equiv
(d\tila^{*}-\tila^{*}\bcQ)Q^{A}
=d \tila^{*}(Q^{A})-\tila^{*}(\{\Theta,Q^{A}\}).
\label{fieldstrengthH}
\end{align}

The gauge transformation of the gauge fields corresponding to the coordinate $z$ is obtained by taking the degree $|z|$ part of the super field strength\footnote{In fact, this formula derives a BRST transformation.
A gauge parameter is a Grassmann odd ghost.} \cite{Ikeda:2012pv},
\begin{align}
\delta Z 
\equiv (\bcd \circ a^{*}(z) - a^{*} \circ \bcQ(z)) \big|_{|z|}.
\label{superfeildstrength}
\end{align}
We introduce a \emph{degree $-1$ map},
$\tiltila: T[1]\Sigma \rightarrow {\cal M}_n$, such that,
for a function of degree $k$ on the target space, 
the pullback $\tila^*$ chooses the component of $(k-1)$-th order in $\theta^\mu$ in the superfield expansion, i.e.,
\begin{align}
\tiltila^{*}(z)& = a^{*}(z) \big|_{|z|-1} 
= \frac{1}{(k-1)!} d\sigma^{\mu_1} \wedge \cdots \wedge d\sigma^{\mu_{k-1}} 
Z^{(k-1)}_{\mu_1 \cdots \mu_{k-1}}(\sigma).
\end{align}

Then, the gauge parameters have degree $|z|-1$, thus, are of ghost number $1$.
We denote the gauge parameters of each transformation 
as
\begin{align}
\epsilon^a & \equiv \tiltila^{*}(q^a), \quad \mu^A \equiv \tiltila^{*}(Q^A) , \\
\epsilon^{\prime}_a & \equiv \tiltila^{*}(p_a) , \quad \mu^{\prime}_A \equiv \tiltila^{*}(P_A). 
\label{gaugetransfparameters}
\end{align}
A gauge transformation is obtained by the following formula,
\begin{align}
\delta \tila^* (z) 
\equiv (d \circ \tiltila^{*}(z) - \tiltila^{*} \circ \bcQ(z)).
\label{superfeildstrength}
\end{align}
The de Rham differential of $F_z$ satisfies
\begin{align}
dF_z=-F\circ\bcQ(z)+\tila^{*}\circ\bcQ^{2}(z).
\end{align}
Thus, from $\bcQ^2=0$, we get the Bianchi identity.


\subsection{Constraints on the conjugate fields 
and residual gauge symmetry
}\label{gaugefixingresidual}
In general, the QP-structure on the QP-manifold ${\cal M}_n = T^*[n]{\cal N}$ induces a substructure of a symplectic Lie $n$-algebra on the non-graded space, $T^*(W \oplus V)$.
It means that the gauge algebra of the pullback of the independent coordinates,
\begin{align}
\tila^{*}(q^{a})& =A^{a},\ \tila^{*}(Q^{A})=B^{A},\\
\tila^{*}(p_{a})& =C_a, \ \tila^{*}(P_{A})=D_A,
\end{align}
is a subalgebra of this symplectic Lie $n$-algebra,
where $C_a$ is an $(n-1)$-form auxiliary gauge field and
$D_A$ is an $(n-2)$-form auxiliary gauge field.

We understand that 
\eqref{generalfieldstrength} gives a field strength with this gauge
symmetry.
Here, we have gauge transformations with four independent gauge parameters \eqref{gaugetransfparameters}.
In order to obtain higher gauge fields and gauge symmetry on the field strength level, we 
impose constraints on the
auxiliary gauge fields $(C_a, D_A)$,
using the extra gauge degrees of freedom.
This reduces the intricate gauge structure to an extension of a differential crossed module. In general, this extension goes beyond the structure of a semistrict Lie $2$-algebra, which brings us into the position to generate interesting higher gauge theories.

The simplest possibility is
to constrain the auxiliary superfields such that $\bcC_a = \bcD_A =0$.
Then, the theory reduces to the one analyzed in the literature \cite{Baez:2010ya}, which is a differential crossed module.
To study covariantization, we consider a nontrivial reduction.

We start with superfields including ghosts and antifields
by using the pullback, $a^{*}(z) = \bcZ(\sigma, \theta)$, of the embedding map, $a: T[1]\Sigma \rightarrow {\cal M}_n$, explained in subsection \ref{HGTfromQP}.
From the degree zero component part, we can read off the field strengths $F^a$ and $H^{A}$ in terms of $A^a = A^{(1)a}$ and $B^A = B^{(2)A}$.
After a restriction of the auxiliary fields, 
there remain gauge symmetries related to $A^a$ and $B^A$.
The residual gauge transformation is on-shell closed by construction.
By choosing a proper restriction, 
we can obtain off-shell covariant field strengths. This procedure is applied in section \ref{hgt5d}.

%% file: sec03.tex
\end{comment}

\section{Hamiltonian functions of higher gauge theories}
\subsection{Hamiltonian functions on the target space ${\cal M}_n$}
As mentioned before, on the manifold ${\cal M}_n$
the Hamiltonian function $\Theta$ is of degree $n+1$.
As we shall see, if the Hamiltonian function is linear in the conjugate coordinates 
$(p_a, P_A)$, it realizes a higher gauge theory with
semistrict Lie 2-algebra structure.
In this paper, we want to consider deformations of this structure.

We start with the most general Hamiltonian function on ${\cal M}_n$
and expand it in
conjugate coordinates $(p_a, P_A)$,
\begin{align}
\Theta &= \sum_{k}
\Theta^{(k)},
\label{expansionsofTheta}
\end{align}
where $\Theta^{(k)}$ is a $k$-th order function in $(p_a, P_A)$.

We distinguish the following cases.
\paragraph{A)}
$n \geq 6$:
Since the degrees of $(p_a, P_A)$ are $(n-1, n-2)$,
the degree of $\Theta^{(k)}$ for $k \geq 2$ is 
larger than $2n-4$.
Therefore, if $n \geq 6$, then $\Theta^{(k)} = 0$ for $k \geq 2$ by 
degree counting, i.e. the general form of the Hamiltonian function is
\begin{align}
\Theta = \Theta^{(0)} + \Theta^{(1)}.
\end{align}
\paragraph{B)}
$n = 4, 5$:
In this case, $\Theta^{(k)} = 0$ for $k \geq 3$ by degree counting.
Therefore, the expansion stops at second order,
\begin{align}
\Theta = \Theta^{(0)} + \Theta^{(1)} + \Theta^{(2)}.
\end{align}
%
%
Only for $n \leq 5$ the Hamiltonian $\Theta$
provides interesting possibilities for deformations.
This case is interesting for physics, 
since it corresponds to a higher gauge theory 
in $5$ and $6$ dimensions.
We discuss the cases $n=4, 5$ in detail in section \ref{hgt5d}
and \ref{hgt6d}. 
\paragraph{C)}
$n=2, 3$:
The Hamiltonian $\Theta$ contains more deformation terms.
%
%
For $n=3$, the local coordinates on the graded manifold ${\cal M}_{3}$ are $(q^{a},Q^{A},p_{a},P_{A})$ with degree $(1,2,2,1)$. 
By taking $U = W \oplus V^*$, 
$T^*[3](W[1] \oplus V[2]) \simeq T^*[3](U[1])$,
the QP-manifold defines a Lie $3$-algebra structure on $U$ \cite{Baez:2003fs}.

In the $n=2$ case,
since $(q^{a},Q^{A}, p_{a},P_{A})$ is of degree $(1, 2, 1, 0)$,
the graded manifold is
${\cal M}_2 =  T^*[2](W[1] \oplus V[2]) 
\simeq  T^*[2]E[1]$, where $E \rightarrow V^*$
is a trivial vector bundle on $V^*$ with fiber $W$.
Since $V^*[0]$ is regarded as base manifold,
we can consider any function of $P_A$ in the Hamiltonian.
Then, this defines a Courant algebroid on $E$ \cite{Liu-Weinstein-Xu, Roytenberg:1999}.

\subsection{Hamiltonian function of the semistrict Lie 2-algebra}\label{semistrictLie2algebra}
First, we show that 
the Hamiltonian function $\Theta^{(1)}$ reproduces a
semistrict Lie $2$-algebra for general $n$.
It contains the following terms,
\begin{align}
\Theta^{(1)}
&= t^{a}_{A}Q^{A}p_{a} +
(-1)^{n}\frac{1}{2}f^{c}_{ab}q^{a}q^{b}p_{c}+\alpha^{B}_{aA}q^{a}Q^{A}P_{B}
+   (-1)^{n}\frac{1}{3!}T^{A}_{abc}q^{a}q^{b}q^{c}P_{A},
\label{eq:bneruibwrubu}
\end{align}
where $t^a_A, f_{ab}^c, \alpha_{aA}^B$ and $T_{abc}^A$
are structure constants.
This function defines a semistrict Lie $2$-algebra,
which is equivalent to a 2-term $L_{\infty}$-algebra
\cite{Baez:2003fs}.

The classical master equation, $\{\Theta^{(1)},\Theta^{(1)}\}=0$,
implies the following conditions on the structure constants,
\begin{align}
\frac{1}{2}f^{d}_{e[a}f^{e}_{bc]}&-\frac{1}{3!}t^{d}_{A}T^{A}_{abc}=0,\\
t^{c}_{A}f^{a}_{cb}&-t^{a}_{B}\alpha^{B}_{bA}=0,\label{Crossedmodule_eq2}\\
\frac{1}{2}\alpha^{B}_{cA}f^{c}_{ab}&+\alpha^{B}_{[a|C|}\alpha^{C}_{b]A}+\frac{1}{2}t^{c}_{A}T^{B}_{cab}=0,\\
\frac{3}{2}f^{e}_{[ab}T^{A}_{cd]e}&+\alpha^{A}_{[a|B|}T^{B}_{bcd]}=0,\\
\alpha^{C}_{a(A}t^{a}_{B)}&=0.
\end{align}
For $T^{A}_{abc} = 0$, these relations reduce to the strict Lie 2-algebra, which is equivalent to the differential crossed module 
\cite{Baez:2010ya}.

The correspondence of the above structure induced by the Hamiltonian 
$\Theta^{(1)}$ and
the semistrict Lie 2-algebra 
is given by
the following bracket and derived brackets,
\begin{align}
[g_1,g_2]&= - \{\{g_1,\Theta^{(1)} \}, g_2\} \big|_{W^*}, 
\label{semistrictoperation1}
\\
\undert(h) &=\{\Theta^{(1)} , h\} \big|_{W^*}, 
\label{semistrictoperation2}
\\
\underalpha(g)h &=\{\{g,\Theta^{(1)} \}, h\}, 
\label{semistrictoperation3}
\\
[g_1,g_2,g_3]&= - \{\{\{g_1,\Theta^{(1)} \}, g_2\}, g_3\},
\label{semistrictoperation4}
\end{align}
where $g_1, g_2, g_3 \in W^*$ and $h \in V^*$.\footnote{We omit the pullback from the shifted vector space to the ordinary vector space for simplicity.}

For details and notation, see appendix A.



\subsection{Canonical transformations on ${\cal M}_n$ }\label{canonicaltransfofn}
In this subsection, we consider canonical transformations on ${\cal M}_n$. 
Two higher gauge theories are equivalent if their defining Hamiltonians can be related by a canonical transformation. 

Let us identify all possible canonical transformations $\alpha$. 
For this, we expand $\alpha$ in conjugate coordinates $(p_a, P_A)$,
\begin{align}
\alpha = \sum_{k} \alpha^{(k)},\label{eq:canonicalbPP}
\end{align}
where $\alpha^{(k)}$ is a $k$-th order function in the
conjugate coordinates.
Since $\alpha^{(k)}$ is of degree $n$, 
by degree counting, we find that
for $n \geq 5$,
$\alpha$ has two terms, $\alpha = \alpha^{(0)} + \alpha^{(1)}$.
For $n = 4$, 
$\alpha^{(2)}$ is nonzero,
i.e. $\alpha = \alpha^{(0)} + \alpha^{(1)} + \alpha^{(2)}$.
For $n =2, 3$, $\alpha$ contains more terms.


In the following, we do not consider $n \leq 3$ further.
Then, the general forms of $\alpha^{(0)}$, $\alpha^{(1)}$ and $\alpha^{(2)}$ are
\begin{align}
\alpha^{(0)} &= \sum_{\mu+2\nu=n, \mu \geq 0, \nu \geq 0}
m^{(\mu,\nu)} 
\nonumber \\ 
&= \sum_{\mu+2\nu=n, \mu \geq 0, \nu \geq 0} \frac{1}{\mu!\nu!} m_{a_{1}\cdots a_{\mu},A_{1}\cdots A_{\nu}} q^{a_{1}}\cdots q^{a_{\mu}}Q^{A_{1}}\cdots Q^{A_\nu},
\\
\alpha^{(1)} &= N + M + \gamma
= N_{A}^{B}Q^{A}P_{B} + M_a^b q^a p_b + 
\frac{1}{2}\gamma^{A}_{ab}q^{a}q^{b}P_{A}, \\
\alpha^{(2)} &= \beta =\frac{1}{2}\beta^{AB}P_{A}P_{B},
\end{align}
where 
$m_{a_{1}\cdots a_{\mu},A_{1}\cdots A_{\nu}}$,
$N_{A}^{B}$, $M_a^b$ and 
$\gamma^{A}_{ab}$ are constants and we defined
$N \equiv N_{A}^{B}Q^{A}P_{B}$, 
$M \equiv M_a^b q^a p_b$ and 
$\gamma \equiv \frac{1}{2}\gamma^{A}_{ab}q^{a}q^{b}P_{A}$.
$\beta^{AB}$ is a symmetric constant.

The authors of \cite{Gruetzmann:2014ica} discuss transformations generated by 
terms corresponding to $\alpha^{(1)}$ above,
as degree preserving coordinate transformations.
In this section, we consider canonical transformations 
of QP-manifolds on $T^{*}[n]{\cal N}$.
Their effect on higher gauge theories will be discussed in the next subsection.
In the following, we investigate the canonical transformations generated by each $\alpha^{(i)}$, respectively.


\paragraph{i)} $\alpha^{(0)}$:
The canonical transformation of a function $f$ of degree $k$ 
by
\begin{align}
m^{(\mu,\nu)}=\frac{1}{\mu!\nu!}m_{a_{1}\cdots a_{\mu},A_{1}\cdots A_{\nu}}q^{a_{1}}\cdots q^{a_{\mu}}Q^{A_{1}}\cdots Q^{A_\nu}. 
\end{align}
The first term is  
\begin{align}
\{f,m^{(\mu,\nu)}\}
=&-(-1)^{k(n-1)}\frac{1}{(\mu-1)!\nu!}\frac{\partial f}{\partial p_{a}}m_{aa_{1}\cdots a_{\mu-1},A_{1}\cdots A_{\nu}}q^{a_{1}}\cdots q^{a_{\mu-1}}Q^{A_{1}}\cdots Q^{A_\nu}\nonumber\\
&-(-1)^{n(k-n)}\frac{1}{\mu!(\nu-1)!}\frac{\partial f}{\partial P_{A}}m_{a_{1}\cdots a_{\mu},AA_{1}\cdots A_{\nu-1}}q^{a_{1}}\cdots q^{a_{\mu}}Q^{A_{1}}\cdots Q^{A_{\nu-1}},
\end{align}
where $\mu+2\nu=n$. 
Investigation of higher order terms is not necessary.
We observe that this class of canonical tranformations decreases the 
order of the conjugate coordinates.
These transformations 
do not change the field strengths 
\eqref{fieldstrengthF} and \eqref{fieldstrengthH}
and the gauge transformations given
in subsection \ref{semistrictLie2algebra}.
Therefore, we do not consider transformations of this type in this paper.

\paragraph{ii)} $\alpha^{(1)}$:
The canonical transformation generated by the term $\gamma=\frac{1}{2}\gamma^{A}_{ab}q^{a}q^{b}P_{A}$ is
\begin{align}
e^{\delta_{\gamma}}p_{a} &= p_{a}+(-1)^{n}\gamma_{ab}^{A}q^{b}_{A},
\label{gammatransfofp}
\\
e^{\delta_{\gamma}}Q^{A} &= Q^{A}+\frac{1}{2}\gamma_{ab}^{A}q^{a}q^{b}.
\label{gammatransfofQ}
\end{align}
This transformation mixes elements of $V$ and $W$.

The canonical transformation generated by $N=N_{A}^{B}Q^{A}P_{B}$
is an automorphism on $V$ and gives an exponential map of the matrix $N_{A}^{B}$,
\begin{align}
e^{\delta_{N}}Q^{A}=(e^{N})_{B}^{A}Q^{B},\\
e^{\delta_{N}}P_{A}=(e^{-N})^{B}_{A}P_{B}.
\end{align}

The canonical transformation 
$M = M_a^b q^a p_b$ generates an automorphism on $W$, similar to the action of $N$
on $V$.

\paragraph{iii)} $\alpha^{(2)}$:
We call this transformation a $\beta$-transformation. It is only possible for $n \leq 4$ and will be used in the discussion of the case $n=4$, below.\\

%% file: sec04.tex
\end{comment}

\section{Higher gauge theories 
in 5 dimensions
}\label{hgt5d}
In the previous sections, we discussed the structure of the Hamiltonian and canonical transformations on ${\cal M}_{n}$. 
Here, we discuss the field theory for the specific case $n=4$, i.e. ${\cal M}_{4} = T^{*}[4]{\cal N}$. 
In this case, $\Theta^{(2)}$ can be included in the Hamiltonian function.

\subsection{General form of the Hamiltonian function and Lie 4-algebras}
In this subsection we describe the structure of the Hamiltonians based on ${\cal M}_{4}$. For this we introduce local coordinates $(q^{a},Q^{A}, p_{a},P_{A})$ of degree $(1,2,3,2)$, respectively.
Since $\Theta$ is of degree 5, the Hamiltonian function is at most a second order function in $(p_a, P^A)$, by degree counting, and can be expanded as
$\Theta = \Theta^{(0)} + \Theta^{(1)} + \Theta^{(2)}$. 
Note that $\Theta^{(2)}$ is nonzero only for $n \leq 5$.
Therefore, the concrete expressions are
\begin{align}
\Theta^{(0)}&=\frac{1}{5!}m_{abcde}q^{a}q^{b}q^{c}q^{d}q^{e}+\frac{1}{3!}m_{abcA}q^{a}q^{b}q^{c}Q^{A}+\frac{1}{2}m_{aAB}q^{a}Q^{A}Q^{B},\label{eq:pescatora}\\
\Theta^{(1)}&=\frac{1}{2}f^{c}_{ab}q^{a}q^{b}p_{c}+t^{a}_{A}Q^{A}p_{a}+\alpha^{B}_{aA}q^{a}Q^{A}P_{B}+\frac{1}{3!}T^{A}_{abc}q^{a}q^{b}q^{c}P_{A},\label{eq:Vongolebianco} \\
\Theta^{(2)}&=s^{aA}p_{a}P_{A}+\frac{1}{2}n_{a}^{AB}q^{a}P_{A}P_{B},\label{eq:Genovese}
\end{align}
with additional structure constants $m_{abcde}$, $m_{abcA}$, $m_{aAB}$, $s^{aA}$ and $n_{a}^{AB}$.
We decompose the classical master equation, $\{\Theta,\Theta\}=0$, by 
degree into 
\begin{align}
&\{\Theta^{(0)},\Theta^{(0)}\}=0,\label{eq:strawberry}\\
&\{\Theta^{(0)},\Theta^{(1)}\}+\{\Theta^{(1)},\Theta^{(0)}\}=0,\label{eq:raspberry}\\
&\{\Theta^{(1)},\Theta^{(1)}\}+\{\Theta^{(0)},\Theta^{(2)}\}+\{\Theta^{(2)},\Theta^{(0)}\}=0,\label{eq:blueberry}\\
&\{\Theta^{(1)},\Theta^{(2)}\}+\{\Theta^{(2)},\Theta^{(1)}\}=0,\label{eq:cranberry}\\
&\{\Theta^{(2)},\Theta^{(2)}\}=0.\label{eq:blackberry}
\end{align}
Concerning the solution of this system of equations we can distinguish four different cases.
\paragraph{I)} Observe that if $\Theta^{(1)}\neq 0$ and $\Theta^{(0)} = \Theta^{(2)} = 0$, then the master equation induces a semistrict Lie 2-algebra structure.
\paragraph{II)} For $\Theta^{(1)}\neq 0$, $\Theta^{(0)} \neq 0$ and $\Theta^{(2)} = 0$, the semistrict Lie 2-algebra structure is not deformed and the induced field strengths as well as the gauge structure are not changed.
\paragraph{III)} In the case, where $\Theta^{(1)}\neq 0$, $\Theta^{(2)} \neq 0$ and $\Theta^{(0)} = 0$, a deformation of the gauge structure as an extension of the semistrict Lie 2-algebra is induced.
\paragraph{IV)} In the most general case ($\Theta^{(i)}\neq 0 \,\, \forall i=0,1,2$), a deformation of the gauge structure as well as the semistrict Lie 2-algebra structure itself is induced. Then, a new type of 2-form gauge theory can be obtained.\\

In this paper, we focus on case III) and analyze extensions of higher gauge structures that avoid the fake curvature condition.\footnote{All relations between the structure constants are listed in 
appendix \ref{Appendixn4}.}
More general cases will be investigated in separate publications.

Let us discuss the resulting classical master equation.
Since there will be additional contributions due to $\Theta^{(2)}$, the conditions including the structure constants $s^{aA}$, $n_a^{AB}$ and $T_{abc}^A$ are
given by
\begin{align}
s^{a(A}n^{BC)}_{a}&=0,
\label{condition6}
\\
s^{cA}f^{b}_{ca}&+\alpha^{A}_{aB}s^{bB}-t^{b}_{B}n^{AB}_{a}=0,
\label{condition7} \\
\frac{1}{2}s^{c(A}T^{B)}_{abc}&+\frac{1}{4}n^{AB}_{c}f^{c}_{ab}+\alpha^{(A}_{[a|C|}n^{B)C}_{b]}=0,
\label{condition8}
\\
s^{a(A}\alpha^{B)}_{aC}&+\frac{1}{2}t^{a}_{C}n^{AB}_{a}=0,
\label{condition9}\\
t^{[a}_{A}s^{b]A}&=0.
\label{condition10}
\end{align}

A QP-structure with $\Theta = \Theta^{(1)} + \Theta^{(2)}$
on ${\cal M}_4 = T^*[4](W[1] \oplus V[2])$
induces the structure of a symplectic Lie $4$-algebra on $T^*(W \oplus V)
\simeq W \oplus V \oplus W^* \oplus V^* \simeq \mathfrak{g}^* 
\oplus \mathfrak{h}^* \oplus \mathfrak{g} \oplus \mathfrak{h}$.

For $g \in W^*$, $h \in V^*$, $\gvee \in W$ and $\hvee \in V$ 
we can introduce symmetric pairings of $W^*$ and $W$, $\bracket{-}{-}_+$,
and antisymmetric pairings of $V^*$ and $V$, $\bracket{-}{-}_-$, 
induced from the P-structure,
\begin{align}
\bracket{g}{\gvee}_+ &\equiv \{g, \gvee \} = \{\gvee, g\}, 
\\
\bracket{h}{\hvee}_- &\equiv \{h, \hvee \} = - \{\hvee, h\}.
\end{align}
In this case, $\{\Theta, \Theta \}=0$ is decomposed into
\begin{align}
& \{\Theta^{(1)}, \Theta^{(1)} \}=0, 
\label{5dmaster1}
\\
& \{\Theta^{(1)}, \Theta^{(2)} \}=0, 
\label{5dmaster2}
\\
& \{\Theta^{(2)}, \Theta^{(2)} \}=0.
\label{5dmaster3}
\end{align}
Equation \eqref{5dmaster1} defines a semistrict Lie 2-algebra structure on 
$W^* \oplus V^*$ as discussed in subsection \ref{semistrictLie2algebra}.
Thus, the above system of equations contains a semistrict Lie 2-algebra $([-,-], [-,-,-], \undert, \underalpha)$ as subalgebra.
Each operation of that subalgebra is defined by 
\eqref{semistrictoperation1}--\eqref{semistrictoperation4}.

Introducing $\Theta^{(2)}$,
we obtain two additional maps
$\unders: W \rightarrow V^*$ and
$\undern: W^* \times V \rightarrow V^*$
corresponding to the new structure constants
by the following graded Poisson bracket and 
derived bracket,
\begin{align}
\unders(\gvee) & = \{\Theta, \gvee \} |_{V^*},
\\
\undern(g)(\hvee) &= - \{ \{g, \Theta\}, \hvee\} \big|_{V^*},
\end{align}
where $g \in W^*$, $\gvee \in W$ and $\hvee \in V$.\footnote{Note that we omit the pullbacks for simplicity.}

It is useful to define the following 
related operations 
$\tildes: W \times V \rightarrow \bC$
and 
$\tilden: W \times V \times V \rightarrow \bC$,
\begin{align}
& \tildes(\gvee, \hvee) = - \{\{\gvee, \Theta \}, \hvee \},
\\
& 
\tilden(\gvee, \hvee_1, \hvee_2) = \{\{\{\gvee, \Theta \}, 
\hvee_1 \}, \hvee_2 \},
\end{align}
as well as 
$\unders^*: V \rightarrow W^*$
and 
$\undern^*: V \times V \rightarrow W$,
\begin{align}
& \unders^*(\hvee) = \{\Theta, \hvee \},
\\
& 
\undern^*(\hvee_1, \hvee_2) = \{\{\hvee_1, \Theta \} \hvee_2\},
\end{align}
such that $\undern^*(\hvee_1, \hvee_2) = \undern^*(\hvee_2, \hvee_1)$, where $\gvee \in W$, $\hvee, \hvee_1, \hvee_2 \in V$.
The above operations are not independent, since
\begin{align}
& \tildes(\gvee, \hvee) = - \bracket{\unders(\gvee)}{\hvee}_-
= \bracket{\unders^*(\hvee)}{\gvee}_+,
\\
& \tilden(g, \hvee_1, \hvee_2) 
= - \bracket{\undern(g)(\hvee_1)}{\hvee_2}_-
= \bracket{\undern^*(\hvee_1, \hvee_2)}{g}_+.
\end{align}
From the conditions \eqref{5dmaster2} and \eqref{5dmaster3}
we obtain the following relations including 
the additional operations $\unders$ and $\undern$,
\begin{align}
& \tildes(\undern^*(\hvee_1, \hvee_2), \hvee_3) 
+ (\hvee_1,\hvee_2, \hvee_3 \ \mbox{symmetric}) = 0,
\label{condition16}
\\
& \bracket{[\unders^*(\hvee), g]}{\gvee}_+
+ \bracket{\alpha(g)\unders(\gvee)}{\hvee}_-
- \bracket{\undert \cdot \undern(g)(\hvee)}{\gvee}_+
=0,
\label{condition17}
\\
& \bracket{[g_1, g_2, \unders^*(\hvee_1)]}{\hvee_2}_-
+ \bracket{\undern([g_1, g_2])(\hvee_1)}{\hvee_2}_-
+ \bracket{\underalpha(g_1)\undern(g_2)(\hvee_1)}{\hvee_2}_-
\nonumber \\
&\qquad
+ (g_1,g_2 \ \mbox{antisymmetric}, \hvee_1,\hvee_2 \ \mbox{symmetric}) = 0,
\label{condition18}
\\
& \bracket{\underalpha \cdot \unders^*(\hvee_1)(h)}{\hvee_2}_-
+ \bracket{\undert(h)}{\undern^*(\hvee_1, \hvee_2)}_+
+ (\hvee_1 \leftrightarrow \hvee_2) = 0,
\label{condition19}
\\
& \bracket{\undert \cdot \unders (\gvee_1)}{\gvee_2}_+
- (\gvee_1 \leftrightarrow \gvee_2) = 0.
\label{condition20}
\end{align}
Using the local coordinate expressions,
\begin{align}
[p_a, p_b]& = f_{ab}^c p_c, \\
\undert(P_A) & = t^a_A p_a, \\
\underalpha(p_a) P_A &= \alpha_{aA}^B P_B, \\
[p_a, p_b, p_c]&= T^A_{abc} P_A, 
\\
\unders(q^a) &= s^{aA} P_A, 
\\
\undern(p_a)(Q^A) &= n_{a}^{AB} P_B,
\end{align}
one shows that the equations \eqref{condition16}--\eqref{condition20} are equivalent to
the equations \eqref{condition6}--\eqref{condition10}.


\subsection{Special solutions of the master equation}
Here, we analyze the relations between the structure constants from the classical master equation for vanishing $\Theta^{(0)}$
and show that there exists a nontrivial solution.
If we take $T_{abc}^A=0$, then the structure constants $\alpha$, $f$ and $t$ in $\Theta^{(1)}$ define a differential crossed module. 
However, there are additional conditions 
on the structure constants $s$ and $n$ in $\Theta^{(2)}$,
given by \eqref{condition6}--\eqref{condition10} with $T_{abc}^A=0$.
In the following, we summarize solutions to the master equation. See appendix \ref{sec:CaseIII} for details on the calculations.

From equation \eqref{condition10}, we can define a symmetric 
constant by $G^{ab} \equiv t^{a}_{A}s^{bA}$. In general, $G^{ab}$ is not invertible. Then, we assume that there exists an invertible metric $g_{ab}$ on $W$,  and we define $s_{a}^{A} \equiv g_{ab}s^{bA}$.
Furthermore, we assume
\begin{align}
s_{a}^{A}t^{a}_{B}=\delta^{A}_{B}.\label{eq:stdelta}
\end{align}
Introducing the matrix ${\cal P}_{b}^{a}=t^{a}_{A}s_{b}^{A}$, we can write $G^{ab}=\mathcal{P}^a_cg^{cb}$, where $g^{ab}$ is the inverse matrix of $g_{ab}$. 
Under the assumption (\ref{eq:stdelta}), ${\cal P}$ is a projector.
Then, \eqref{condition7} becomes
\begin{align}
n^{AB}_{a}=s^{B}_{b}s^{cA}f^{b}_{ca}+s^{B}_{b}s^{bC}\alpha^{A}_{aC}.
\end{align}
The crossed module relation (\ref{Crossedmodule_eq2})
gives $\alpha^{A}_{aB}$ and $n_a^{AB}$ as
\begin{align}
\alpha^{A}_{aB} &=s^{A}_{b}t_{B}^{c} f^{b}_{ca}, \label{ALPHA}
\\
n^{AB}_{a} &=2s^{c(A}s^{B)}_{b}f^{b}_{ca}.
\end{align}
From \eqref{condition8}, we obtain a condition on the structure of $f$,
\begin{align}
g^{dg}s^{(A}_{g}\{s^{B)}_{e}(\delta^{f}_{c}-{\cal P}^{f}_{c})
f^{e}_{f[a}f^{c}_{b]d}\}=0,
\label{eq:beiorguirsbijlzvjn}
\end{align}
while the other conditions are satisfied automatically.
The explicit form of the total Hamiltonian related to this solution of the classical master equation is given by
\begin{align}
\Theta
&=\frac{1}{2}f^{c}_{ab}q^{a}q^{b}p_{c}+t^{a}_{A}Q^{A}p_{a}+\alpha^{B}_{aA}q^{a}Q^{A}P_{B}+s^{aA}p_{a}P_{A}+s^{cA}s^{B}_{b}f^{b}_{ca}q^{a}P_{A}P_{B}. \label{THETAI}
\end{align}
In the special case, where $\calP^a_b = \delta^a_b$, we find that $G^{ab}$ is invertible. Then, \eqref{eq:beiorguirsbijlzvjn} is automatically satisfied and \eqref{ALPHA} implies
\begin{align}
s^{B}_{(a}\alpha_{b)B}^{A}=0.
\end{align}
Finally, in this special case, \eqref{THETAI} reduces to
\begin{align}
\Theta
&=\frac{1}{2}f^{c}_{ab}q^{a}q^{b}p_{c}+t^{a}_{A}Q^{A}p_{a}+\alpha^{B}_{aA}q^{a}Q^{A}P_{B}+s^{aA}p_{a}P_{A}+s_{b}^{C}s^{bB}\alpha^{A}_{aC}q^{a}P_{A}P_{B}. \label{THETAII}
\end{align}

\paragraph{Canonical transformations on ${\cal M}_{4}$}
Let us consider the canonical transformations $e^{\delta_{\alpha}}$ on ${\cal M}_{4}$, 
where the general form of the generator $\alpha$ (\ref{eq:canonicalbPP}) contains 
the term $\alpha^{(2)}=\frac{1}{2}\beta^{AB}P_{A}P_{B}$, that we call $\beta$-transformation.
The Hamiltonians \eqref{THETAI} and \eqref{THETAII} can be generated by $\beta$-transformation from the differential crossed module. In general, we have ${\cal P}^{a}_{b}s^{b}_{A}=s^{a}_{A}$. Twist by the canonical transformation
\begin{equation}
\beta^{AB}=s^{aA}s_{a}^{B}=g^{ab}s_{a}^{A}s_{b}^{B},
\end{equation}
we find
\begin{equation}
\Theta = e^{\delta_{\beta}} \Theta^{(1)},
\end{equation}
for the Hamiltonian functions \eqref{THETAI} and \eqref{THETAII} by using the respective solution of the classical master equation.
Thus, we understand that this set of special solutions to 
the master equation exhibits the structure of a differential crossed module.
However, we will show in the following subsections, that
one can circumvent the fake curvature condition, usually
related to models based on the crossed model,
by reducing the gauge freedom of the auxiliary gauge fields.

%% file: sec04-2.tex
\end{comment}

\subsection{Constraints on the conjugate fields
}
Based on the general theory explained in subsection \ref{gaugefixingresidual},
we consider the restriction of the 5-dimensional theory.

The pullback $a^*$ maps the four coordinates to superfields as follows,
\begin{align}
\bcA^{a}& \equiv a^{*}(q^{a}),\ \bcB^{A} \equiv a^{*}(Q^{A}),\\
\bcC_a & \equiv a^{*}(p_{a}), \ \bcD_A \equiv a^{*}(P_{A}),
\end{align}
where $(\bcA, \bcB, \bcC, \bcD)$ are of degree $(1, 2, 3, 2)$.
The QP-manifold structure on ${\cal M}_4$ induces the structure of a symplectic Lie $4$-algebra on $T^*(W \oplus V)$. 
The superfields inherit this structure as gauge symmetry.
The super field strengths are given by
\begin{align}
\bcF^{a}&= \bcd \bcA^{a} -\frac{1}{2}f^{a}_{bc} \bcA^{b} \bcA^{c}
- t^{a}_{A} \bcB^{A} - s^{aA} \bcD_{A},
\label{superfieldstrength5d1}
\\
\bcH^{A}&= \bcd \bcB^{A} + \alpha_{aB}^{A} \bcA^{a} \bcB^{B}
+ \frac{1}{3!} T^{A}_{abc} \bcA^{a} \bcA^{b} \bcA^{c}
+ s^{bA} \bcC_b + n^{AB}_a \bcA^a \bcD_{B},
\label{superfieldstrength5d2}
\\
\bcF^{(C)}_{a}&= \bcd \bcC_{a} - f^{c}_{ab} \bcA^{b} \bcC_{c}
- \alpha_{aB}^{A} \bcB^{B} \bcD_{A}
- \frac{1}{2} T^{A}_{abc} \bcA^{b} \bcA^{c} \bcD_A
- \frac{1}{2}n^{AB}_a \bcD_A \bcD_B,
\label{superfieldstrength5d3}
\\
\bcF^{(D)}_{A}&=\bcd \bcD_{A} - t^{a}_{A} \bcC_a 
- \alpha_{aA}^{B} \bcA^{a} \bcD_{B},
\label{superfieldstrength5d4}
\end{align}
where $\bcF^{(C)}$ and $\bcF^{(D)}$ are the super field strengths of 
$\bcC$ and $\bcD$, respectively.
When we substitute the component expansions to \eqref{superfieldstrength5d1}--\eqref{superfieldstrength5d1}, then the
corresponding degree $|z|+1$ parts are the field strengths:
\begin{align}
F^{a}&= d A^{a} -\frac{1}{2}f^{a}_{bc} A^{b} \wedge A^{c}
- t^{a}_{A} B^{A} - s^{aA} D_{A},
\label{fieldstrength5d1}
\\
H^{A}&= d B^{A} + \alpha_{aB}^{A} A^{a} \wedge B^{B}
+ \frac{1}{3!} T^{A}_{abc} A^{a} \wedge A^{b} \wedge A^{c}
+ s^{bA} C_b + n^{AB}_a A^a \wedge D_{B},
\label{fieldstrength5d2}
\\
F^{(C)}_{a}&= d C_{a} - f^{c}_{ab} A^{b} \wedge C_{c}
- \alpha_{aB}^{A} B^{B} \wedge D_{A}
- \frac{1}{2} T^{A}_{abc} A^{b} \wedge A^{c} \wedge D_A
- \frac{1}{2}n^{AB}_a D_A \wedge D_B,
\label{fieldstrength5d3}
\\
F^{(D)}_{A}&= d D_{A} - t^{a}_{A} C_a 
- \alpha_{aA}^{B} A^{a} \wedge D_{B}.
\label{fieldstrength5d4}
\end{align} 
The degree $|z|$ parts of the component expansions of the super field strengths yield the
gauge transformations,
\begin{align}
\delta A^{a}&= d \epsilon^{a} - f^{a}_{bc} A^{b} \epsilon^{c}
- t^{a}_{A} \mu^{A} - s^{aA} \mu^{\prime}_{A},
\\
\delta B^{A}&= d \mu^{A} + \alpha_{aB}^{A} (A^{a} \wedge \mu^{B}
+ \epsilon^{a} B^{B})
+ \frac{1}{2} T^{A}_{abc} A^{a} \wedge A^{b} \epsilon^{c}
+ s^{bA} \epsilon^{\prime}_b 
\nonumber \\
& + n^{AB}_a (A^a \wedge \mu^{\prime}_{B}
+ \epsilon^a \wedge D_{B}),
\\
\delta C_{a}&= d \epsilon^{\prime}_{a} 
- f^{c}_{ab} (A^{b} \wedge \epsilon^{\prime}_{c} + \epsilon^{b} \wedge C_{c})
- \alpha_{aB}^{A} (B^{B} \wedge \mu^{\prime}_{A} + \mu^{B} \wedge D_{A})
\nonumber \\ & 
- \frac{1}{2} T^{A}_{abc} (2 A^{b} \wedge D_A \epsilon^{c} 
+ A^{b} \wedge A^{c} \wedge \mu^{\prime}_A) - n^{AB}_a D_A \wedge \mu^{\prime}_B, \label{DeltaC}
\\
\delta D_{A}&= d \mu^{\prime}_{A} - t^{a}_{A} \epsilon^{\prime}_a 
- \alpha_{aA}^{B} (A^{a} \wedge \mu^{\prime}_{B} + \epsilon^{a} D_{B}).
\end{align}
The gauge transformations of the higher gauge field strengths are
\begin{align}
\delta F^{a} &= f^{a}_{bc} F^{b} \epsilon^{c},
\\
\delta H^{A} &= \alpha_{aB}^{A} H^{B} \epsilon^a
+ n^{AB}_a F^{(D)}_B \epsilon^a
+ T^{A}_{abc} A^{a} \wedge F^{c} \epsilon^{b}
\nonumber \\
& - \alpha_{aB}^{A} F^a \wedge \mu^{B} 
- n^{AB}_a F^a \wedge \mu^{\prime}_{B}.
\end{align}
We look for nontrivial extensions of the crossed module inside a symplectic Lie $4$-algebra, that lead to off-shell covariant gauge structures. Such an extension can be 
obtained by imposing a constraint on the gauge fields $(C_a, D_A)$.

One nontrivial choice is given by 
\begin{align}
C_a= - K_{abc} F^{b}\wedge A^{c},
\quad
D_A = 0.
\label{reductionconditionCD}
\end{align}
Then, we obtain the field strengths $F^a$ and $H^{A}$ 
in terms of $A^a$ and $B^A$,
\begin{align}
F^{a}&= dA^{a}-\frac{1}{2}f^{a}_{bc}A ^{b}\wedge A ^{c}-t^{a}_{A} B^{A},\\
H^{A}&= d B^{A} + \alpha_{aB}^{A} A^{a} \wedge B^{B}
+ \frac{1}{3!} T^{A}_{abc} A^{a} \wedge A^{b} \wedge A^{c}
- s^{bA} K_{bcd} F^{c} \wedge A^{d}.
\label{gaugefixedH}
\end{align}


In general, the original gauge transformations of the fields 
$(A^a, B^A)$ transform the constraint equations \eqref{reductionconditionCD}.
However, there exist compensating gauge transformations of the fields 
$(C_a, D_A)$ such that 
the conditions \eqref{reductionconditionCD} remain satisfied.

For the special case given in equation \eqref{reductionconditionCD}, where $T_{abc}^A = 0$, 
\begin{align}
K_{abc} = g_{ad}t^{d}_{A}\alpha^{A}_{bB} s_{c}^{B} 
\end{align}
%
we obtain the gauge fixed field strengths of $A^a$ and $B^A$ decoupled from the $\bcC_a$ and $\bcD_A$ components,
\begin{align}
F^{a}&= dA^{a}-\frac{1}{2}f^{a}_{bc}A ^{b}\wedge A ^{c}-t^{a}_{A} B^{A},\\
H^{A}&= dB^{A}+\alpha_{aB}^{A}A^{a}\wedge B^{B}-\alpha_{aB}^{A} s^{B}_{c} F^{a}\wedge A^{c},
\label{gaugefixedH}
\end{align}
which is of the same form as the field strengths given in \cite{Ho:2012nt}. The detailed relations between our formulation and the results given in \cite{Ho:2012nt} will be discussed below.

\subsection{Off-shell covariantization}

In this subsection, we show that we can off-shell covariantize the 3-form curvature in the setting, where the underlying structure is a semi-direct product $W^* = \mathfrak{g} = K\ltimes \mathfrak{h}$. $K$ is a Lie algebra and $\rho$ is a representation of $K$ on the vector space $V^* = \mathfrak{h}$. This setting is in accordance with the special set of solutions that we discussed in the previous subsection. 

The commutator on $\mathfrak{g}$ is defined as
\begin{align}
	[(x,y),(x',y')]=([x,x'],\rho(x)y'-\rho(x')y),
\end{align}
where $x,x'\in K$ and $y,y'\in \mathfrak{h}$. Furthermore, we define the maps $\underline{\alpha}: \mathfrak{g}\rightarrow \text{Der}(\mathfrak{h})$, $\underline{t}: \mathfrak{h}\rightarrow \mathfrak{g}$ and $\underline{s}: \mathfrak{g}\rightarrow \mathfrak{h}$ by
\begin{align}
	\underline{\alpha}((x,y))y' &= \rho(x)y', \\
	\underline{t}(y) &= (0,My), \\
	\underline{s}(x,y) &= M^{-1}y,
\end{align}
where $M$ is an invertible matrix. 
This setting has been used in \cite{Ho:2012nt} in order to construct an off-shell covariant higher gauge theory. We use the index convention $g_a = (g_i, g_A)\in K\ltimes \mathfrak{h}$.
In order to discuss the covariantization of the 3-form curvature $H^A$, we start with an analysis of its gauge transformation,
\begin{align}
	\delta H^A &= \alpha_{aB}^A H^B \epsilon^a - \alpha_{aB}^A F^a \wedge \mu^B - n_a^{AB} F^a \wedge \mu'_{B} + n_{a}^{AB}F^{(D)}_B \epsilon^a \\
	&\equiv \alpha_{aB}^A H^B \epsilon^a - \triangle^A,
\end{align}
where we took $T^A_{abc}=0$. 
We can decompose
\begin{align}
	\triangle^A &= s^A_a t^b_B f^a_{jb}F^j \wedge \mu^B + (s^{bA}s^B_a f^a_{jb} + s^{bB} s_a^A f^a_{jb})F^j \wedge \mu'_B \notag \\
	&\qquad  - n^{AB}_j dD_B \epsilon^j - n^{AB}_j t^a_B C_a \epsilon^j - n^{AB}_j \alpha_{aB}^C A^a \wedge D_C \epsilon^j.
\end{align}
For covariantization we make use of the freedom of the conjugate auxiliary fields $C_a$ and $D_A$. The constraint \eqref{reductionconditionCD} leads to
\begin{equation}
	\triangle^A = s^A_a t^b_B f^a_{jb}F^j \wedge \mu^B + (s^{bA}s^B_a f^a_{jb} + s^{bB} s_a^A f^a_{jb})F^j \wedge \mu'_B - n^{AB}_j t^a_A C_a \epsilon^j.
\end{equation}
We show in the following, that $\triangle^A$ vanishes, if the field configuration is  restricted to a hypersurface determined by the constraint. First, we introduce the gauge parameters $\he^a$ and $\hm^A$ corresponding to the remaining gauge symmetry and require, that the reduced gauge transformation of the one-form gauge field is given by
\begin{equation}
	\delta A^a = D_{0}\epsilon^a - t^a_A \mu^A - s^{aA}\mu'_{A} \equiv D_{0}\he^a - t^a_A\hm^A, \label{ReducedA}
\end{equation}
where we introduced $D_0 \he^a \equiv d\he^a - f_{bc}^a A^b \he^c$. Through application of the projector $(1 - \calP)$ to \eqref{ReducedA} we find
\begin{equation}
	(1-\calP)D_{0}\epsilon = (1-\calP)D_{0}\he.
\end{equation}
Making use of the equation
\begin{align}
	f^a_{ba'}\calP^{a'}_d = \calP^a_c f^c_{ba'}\calP^{a'}_d,
\end{align}
which can be derived from \eqref{eq:beiorguirsbijlzvjn}, we find
\begin{equation}
	\epsilon^a = \he^a +  (\calP \lambda(\he))^a \label{PXOffset}
\end{equation}
for an arbitrary function $\lambda(\he)$. $\lambda(\he)$ has to be of order one in the gauge parameter $\he$ or zero.
Application of $\calP$ to \eqref{ReducedA} leads to
\begin{equation}
	s^{aA}\mu'_{A} + t^a_A\mu^A = t^a_A\hm^A + \calP^a_b D_0 \calP^b_c \lambda(\he)^c. \label{MuHat}
\end{equation}
In the next step, we solve the first constraint $\delta D_A = 0$, which gives
\begin{equation}
	\calP^b_a\epsilon'_b = \calP^b_a[d(s_b^A\mu'_A) - f_{bc}^d A^c (s_d^A\mu'_A)] \equiv \calP^b_a D_0 s_b^B\mu'_B. \label{GaugeFixD}
\end{equation}
Application of $\calP D_0$ to \eqref{GaugeFixD} gives
\begin{equation}
	\calP^{aa'} D_0 \calP^b_{a'} \epsilon'_b = \calP^{ab}F^i f_{ib}^{b'} s_{b'}^B \mu'_B. \label{GaugeFixD2}
\end{equation}
Let us now investigate the covariance of the second constaint equation,
\begin{equation}
        \delta C_a= - g_{ad}t^d_A \alpha_{b' B}^A s^B_c \delta(F^{b'}\wedge A^c). \label{CovSecondGFE}
\end{equation} 
Using the covariance condition following from the first constraint \eqref{GaugeFixD2}, we derive the following condition on $\mu'$ in terms of $\he$ and $\hm$ from the projected part of \eqref{CovSecondGFE}
\begin{equation}
	\calP^{ab} F^i f^{b'}_{ib} s_{b'}^B \mu'_B = \epsilon^j t^a_A t^c_B n^{AB}_j C_c + f^a_{jb} F^j \wedge t^b_B(s_d^B D_0 \calP^d_c \he^c - \hm^B). \label{CovH}
\end{equation}
On the other hand, we can rewrite $\triangle^A$ by
\begin{equation}
	\triangle^A = s^A_a f_{jb}^a F^j\wedge t^b_B(\hm^B + s_d^B D_0 \calP^d_c \lambda(\he)^c) + s^{bA} s^B_a f_{jb}^a F^j\wedge \mu'_B - n_j^{AB} t^b_B C_b \epsilon^j.
\end{equation}
We find, that $\triangle^A = 0$, if $\lambda(\he)^a = -\he^a$ on the hypersurface determined by the two constraints. The remaining condition coming from the orthogonal projection of the second constraint equation
\begin{equation}
	(1-\calP)^b_a\delta C_b = - (1 - \calP)^b_a g_{bd}t^d_A \alpha_{b' B}^A s^B_c \delta(F^{b'}\wedge A^c) = 0,
\end{equation}
imposes a restriction on $(1 - \calP)^a_i\epsilon'_a = \epsilon'_i$.

Finally, the gauge transformation of the two-form gauge field on the gauge-hypersurface is derived to be
\begin{equation}
	\delta B^A = d\hm^A + \alpha_{jB}^A(A^j\wedge \hm^B + \he^j B^B) - \alpha_{jB}^A s^B_c \he^c F^j,
\end{equation}
by using \eqref{PXOffset}, \eqref{MuHat} and \eqref{GaugeFixD}.
Therefore, we showed that after imposing proper constraints, the field strengths transform covariantly under the residual gauge transformations without fake curvature condition.

Let us summarize the form of the fields and their transformation properties on the gauge-hypersurface,
\begin{align}
	F^{a}&= dA^{a}-\frac{1}{2}f^{a}_{bc}A ^{b}\wedge A ^{c}-t^{a}_{A} B^{A},\\
	H^{A}&= dB^{A}+\alpha_{aB}^{A}A^{a}\wedge B^{B}-\alpha_{aB}^{A} s^{B}_{c} F^{a}\wedge A^{c}, \\
	\dha A^a &= d\he^a - f_{bc}^a A^b \he^c - t^a_A\hm^A, \\
	\dha B^A &= d\hm^A + \alpha_{jB}^A(A^j\wedge \hm^B + \he^j B^B) 
- \alpha_{jB}^A s^B_c \he^c F^j, \\
	\dha F^a &=  f^{a}_{bc} F^{b} (\he^c - (\calP \he)^c), \\
	\dha H^A &= \alpha_{aB}^A H^B (\he^a - (\calP \he)^a),
\end{align}
where we introduced $\dha$ symbolizing the reduced gauge transformation $\delta \Phi\big|_{\text{constraint}} = \dha \Phi$ for any field $\Phi$, which means that the diagram
\begin{eqnarray}
\begin{CD}
\Phi @>\delta  >> \delta \Phi \\
@V\text{constraint}VV @VV\text{constraint}V \\
\Phi @>\dha>> \dha \Phi
\end{CD}
\nonumber
\end{eqnarray}
commutes for any field $\Phi$. 

Next, we discuss the closure of the gauge symmetry algebra. For this, we write the gauge transformation as 
\begin{align}
	\tilde\delta A^a &= d\tilde{\epsilon}^a - f_{bc}^a A^b \tilde{\epsilon}^c + t^a_A\tilde{\mu}^A, \\
	\tilde{\delta} B^A &= d\tilde{\mu}^A + \alpha_{jB}^A(A^j\wedge \tilde{\mu}^B - \tilde{\epsilon}^j B^B) 
+ \alpha_{jB}^A s^B_c \tilde{\epsilon}^c F^j,
\end{align}
where the gauge parameters $\tilde{\epsilon}^a$ and $\tilde{\mu}^A$ are ordinary functions.
We find, that two gauge transformations $\tilde{\delta}_1$ and $\tilde{\delta}_2$ close to  $\tilde{\delta}_3$ by $[\tilde{\delta}_1, \tilde{\delta}_2] = \tilde{\delta}_3$ with $\tilde{\epsilon}_3^a = - f^a_{bc} \tilde{\epsilon}_1^b \tilde{\epsilon}_2^c$ and $\tilde{\mu}_3^A = \alpha^A_{bB} (\tilde{\epsilon}_1^b \tilde{\mu}_2^B - \tilde{\epsilon}_2^b \tilde{\mu}_1^B)$, where $\tilde{\delta}_i$ denotes the gauge transformation with respective gauge parameters 
$(\tilde{\epsilon}_i, \tilde{\mu}_i)$. More concretely, we derive
\begin{align}
	[\tilde{\delta}_1, \tilde{\delta}_2] A^a & = d \tilde{\epsilon}_3^a - f_{bc}^a A^b \tilde{\epsilon}_3^c 
+ t^a_A\tilde{\mu}_3^A, \\
	[\tilde{\delta}_1, \tilde{\delta}_2] B^A & =  d\tilde{\mu}_3 ^A + \alpha_{jB}^A(A^j\wedge \tilde{\mu}_3^B - \tilde{\epsilon}_3^j B^B) + \alpha_{jB}^A s^B_c \tilde{\epsilon}_3^c F^j + \Lambda^A,
\end{align}
where
\begin{equation}
	\Lambda^A = \alpha_{jB}^A f^j_{ke} s^B_c \calP^e_b 
(\tilde{\epsilon}_1^b \tilde{\epsilon}_2^c 
- \tilde{\epsilon}_2^b \tilde{\epsilon}_1^c  ) F^k.
\end{equation}
The gauge transformation of $A^a$ is off-shell closed.
Off-shell closure of the gauge transformation of $B^A$ requires
\begin{equation}
\alpha_{jB}^A f^j_{ke} s^B_c \calP^e_b = \alpha^{A}_{jB}f^{j}_{ke}s^{B}_{c} 
t^{e}_{D} s^{D}_{b}=0,
\label{offshellclosednesscondtion3}
\end{equation}
which is satisfied in our example.  In general, it is sufficient if a gauge algebra is closed on-shell, i.e., up to equations of motion.
However, with condition \eqref{offshellclosednesscondtion3} the gauge algebra is closed without 
fake curvature condition.

The field strengths and gauge transformations, that we derived, are of a similar form compared to the ones analyzed in \cite{Ho:2012nt}. To make this similarity more concrete, we provide a brief discussion of the model proposed in \cite{Ho:2012nt} in the following subsection.

\subsection{The Ho-Matsuo model}
The authors of \cite{Ho:2012nt} constructed a covariant 3-form field strength which circumvents the fake curvature condition. Here, we give a brief review of the algebra, gauge transformations and field strengths constructed in \cite{Ho:2012nt}, for comparison.
Let $(W^*, V^*, \undert, \underalpha)$ be a differential crossed module with the additional map $\undershm: W^* \rightarrow V^*$ and the following consistency conditions,
\begin{align}
& \underalpha(g)(\undershm(g^{\prime})) - \underalpha(g^{\prime})(\undershm(g)) 
= \undershm([g, g^{\prime}]),
\label{HMalgebra1} \\
& \underalpha(g)((1 - \undershm \cdot \undert)(h)) =0,
\label{HMalgebra2} 
\\
& \underalpha([g, \undert \cdot \undershm(g^{\prime})])(\undershm(g^{\prime\prime})) =0,
\label{HMalgebra3}
\end{align}
where $g, g^{\prime}, g^{\prime\prime} \in W^*$
and $h, h^{\prime} \in V^*$.
Representing the map $\undershm$ by
\begin{align}
& \undershm(g_a) = s_a^B h_B,
\end{align}
the conditions \eqref{HMalgebra1}--\eqref{HMalgebra3} become 
\begin{align}
\alpha^{A}_{aB}s^{B}_{b}-\alpha^{A}_{bB}s^{B}_{a}+s^{A}_{c}f^{c}_{ab}=0,
\label{eq:check1}\\
\alpha^{A}_{aD}s^{D}_{b}t^{b}_{B}=\alpha^{A}_{aB},\label{eq:check2}\\
s^{B}_{c}\alpha^{A}_{dB}f^{d}_{ae}t^{e}_{D}s^{D}_{b}=0.\label{eq:check3}
\end{align}
The equations \eqref{eq:check1}--\eqref{eq:check2} are satisfied in the Lie 4-algebra model constructed in the previous subsection.
Equation \eqref{eq:check3} is required by the closure of the gauge transformation
of the 2-form gauge field \eqref{offshellclosednesscondtion3}.
The field strengths of the 1- and 2-form fields are defined as follows,
\begin{align}
F_{HM}^{a} &=dA^{a}-\frac{1}{2}f^{a}_{bc}A ^{b}\wedge A ^{c}-t^{a}_{A} B^{A},\\
H_{HM}^{A} &=dB^{A}+\alpha_{aB}^{A}A^{a}\wedge B^{B}-\alpha_{aB}^{A} s^{B}_{c} F_{HM}^{a}\wedge A^{c}.
\end{align}
The gauge transformations of the 1-form and 2-form  
gauge fields are
defined by
\begin{align}
\delta_{HM} A^{a}&= d\he^{a}-f^{a}_{bc}A^{b}\he^{c}+t^{a}_{A}\hm^{A},\\
\delta_{HM} B^{A}&= d\hm^{A}+\alpha^{A}_{aB}A^{a}\wedge \hm^{B}-\alpha_{aB}^{A}\he^{a}B^{B}+\alpha^{A}_{dB}s^{B}_{a}\he^{a}F_{HM}^{d}. \label{deltaBHM}
\end{align}

%% file: sec05.tex
\end{comment}

\section{Master equation in 6 dimensions}\label{hgt6d}
In this section we discuss the field theory on the graded manifold 
${\cal M}_{5}=T^{*}[5](W[1]\oplus V[2])$ with coordinates
$(q^{a},Q^{A}, p_{a},P_{A})$ of degree $(1, 2, 4, 3)$.
The most general form of the Hamiltonian function is given by $\Theta = \Theta^{(0)} + \Theta^{(1)} + \Theta^{(2)}$, where
\begin{align}
\Theta^{(0)}&=\frac{1}{6!}m_{abcdef}q^{a}q^{b}q^{c}q^{d}q^{e}q^{f}+\frac{1}{4!}m_{abcdA}q^{a}q^{b}q^{c}q^{d}Q^{A}\nonumber\\
&+\frac{1}{4}m_{abAB}q^{a}q^{b}Q^{A}Q^{B}+\frac{1}{3!}m_{ABC}Q^{A}Q^{B}Q^{C},\label{eq:pescatora5}\\
\Theta^{(1)}&=-\frac{1}{2}f^{c}_{ab}q^{a}q^{b}p_{c}+t^{a}_{A}Q^{A}p_{a}+\alpha^{B}_{aA}q^{a}Q^{A}P_{B}-\frac{1}{3!}T^{A}_{abc}q^{a}q^{b}q^{c}P_{A},\label{eq:Vongolebianco5} \\
\Theta^{(2)}&=\frac{1}{2}u^{AB}P_{A}P_{B}.\label{eq:Genovese5}
\end{align}
Here, again we focus on the case $\Theta^{(0)} = 0$, which means that we have to consider the following equation: $\{\Theta^{(2)},\Theta^{(1)}\}+\{\Theta^{(1)},\Theta^{(2)}\}=0$. Since $\Theta^{(2)}$ depends only on $P$, we immediately get
\begin{align}
\frac{1}{2}&(\{\Theta^{(1)},\Theta^{(2)}\}+\{\Theta^{(2)},\Theta^{(1)}\})=t^{a}_{A}u^{AB}p_{a}P_{B}+\alpha^{B}_{aA}u^{AC}q^{a}P_{B}P_{C}.
\end{align}
This leads to the following conditions for the antisymmetric bilinear $u^{AB}$:
\begin{align}
t^{a}_{A}u^{AB}=0,\label{eq:kerasa}\ 
\alpha^{[B}_{aA}u^{C]A}=0.
\end{align}
Since the structure constant $s^{aA}$ does not appear in the algebra and
there are no possibilities to induce 
a nonzero tensor of type $s^{aA}$ from the other structure constants, 
we cannot formulate a model analogous to the one we constructed in 5 dimensions.\footnote{Using \eqref{Crossedmodule_eq2} and \eqref{eq:kerasa} one can show that $s^{aA} = 0$ for any tensor of this index structure.}

%% file: discussion.tex
\section{Discussion}
In this paper, we analyzed extensions of higher gauge theories based on a semistrict Lie $2$-algebra. We made use of the QP-manifold description of symplectic Lie $n$-algebras 
and constructed an off-shell covariant higher gauge theory. 
The gauge fields induced by the Lie $n$-algebra inherit its structure as gauge symmetry.
In order to obtain an off-shell covariantized higher gauge theory which circumvents the fake curvature condition, we restrict the auxiliary gauge field configuration to an appropriate hypersurface. The restricted gauge algebra has the structure of an extension of a (semistrict) Lie $2$-algebra.
We analyzed the structure of the QP-manifold 
$T^*[n](W[1] \oplus V[2])$, the general structure of its possible Hamiltonians and its canonical transformations.
It turned out that for $n \geq 6$, i.e., for a theory in 7 dimensions or 
higher, we only obtain a semistrict higher gauge theory.
For $n \leq 5$, i.e., for a theory in 6 dimensions or lower, 
there is a freedom to introduce terms into the Hamiltonian function,
which change the field strengths nontrivially.

In this paper, we analyzed possible deformations by $\Theta^{(2)}$. This is
only possible in dimensions less than 7. We examined the 5 dimensional theory in detail.
Still in this case, there are many choices for imposing conditions on the auxiliary gauge fields. 
We concentrated on the case, where $\mathfrak{g} = K\ltimes \mathfrak{h}$ and $\mathfrak{h}= V^*$, where $K$ is a Lie algebra and $\rho$ is a representation of $K$ on $V^*$. Then, we showed, that by the present method, we can obtain a nontrivial off-shell covariant theory.
In this case, the theory is the same covariantization as the one given in \cite{Ho:2012nt}.
Although this theory, which we constructed, exhibits abelian higher gauge structure, we think that depending on the reduction procedure also nonabelian solutions can be found.

As we discussed in the beginning of section 4, there are several directions to develop the approach given in this paper.
One way is to include the $\Theta^{(0)}$ term. This works in any dimension and, 
 in general, does not change the field strengths $F$ and $H$.
However, it changes the structures of the algebra.
Another possibility is to include algebroid structures, which introduces scalar fields in the theory. 

It is interesting that we could obtain a 2-form gauge theory by reduction of a
Lie $n$-algebra structure. This was performed by imposing constraints on the auxiliary gauge fields on the field theory level. There is also a possibility to interpret this reduction process as gauge fixing of auxiliary gauge fields. 


\section*{Acknowledgments}
The authors would like to thank T. Asakawa, B. Jur\v{c}o, Y. Maeda, V. Mathai, Y. Matsuo, 
H. Muraki, J.-S. Park, J.-H. Park, P. Ritter, C. Saemann, T. Strobl and M. Zabzine 
for interesting lectures, stimulating discussions and valuable comments during the Tohoku Forum for Creativity thematic program "Fundamental Problems in Quantum Physics: Strings, Black Holes and Quantum Information" held in 2015.
 
M.A.H. is supported by Japanese Government (MONBUKAGAKUSHO) Scholarship.
Y.K. is supported by Tohoku University Division for Interdisciplinary Advanced
Research and Education (DIARE).
N.I. and S.W. are supported by the Japan-Belgium Bilateral Joint Research Project of JSPS.

%% file: app01.tex
\end{comment}

\section{Differential crossed modules and semistrict Lie 2-algebras}\label{crossedmodulesemistrictlie2}
First, we briefly explain the crossed module and differential crossed module \cite{Baez:2002jn, Baez:2003fs}.

A \emph{crossed module} is a pair of Lie groups $G$ and $H$
with homomorphisms $t: H \rightarrow G$ 
and $\alpha: G \rightarrow \mathrm{Aut}(H)$
satisfying
\begin{align}
\alpha(t(h))(h^{\prime})&= h h^{\prime} h, \\
t(\alpha(g)h)&=g t(h) g^{-1},
\label{crossedmodule}
\end{align}
for all $g \in G$ and $h, h^{\prime} \in H$.

Let $\mathfrak{g}=\text{Lie}(G)$ and $\mathfrak{h}=\text{Lie}(H)$
be the associated Lie algebras.
The infinitesimal object corresponding to the crossed module is called 
\emph{differential crossed module}. It is a pair of Lie
algebras $\mathfrak{g}$ and $\mathfrak{h}$
with homomorphisms 
$\undert: \mathfrak{h} \rightarrow \mathfrak{g}$
and $\underalpha: \mathfrak{g} \rightarrow \mathrm{Der}(\mathfrak{h})$.
The differentials of the corresponding maps are underlined.
$\undert$ and $\underalpha$ satisfy
\begin{align}
\undert(\underalpha(g)h)&=[g, \undert(h)], \label{differentialcrossedmodule} \\
\underalpha(\undert(h))(h^{\prime})&= [h, h^{\prime}],
\end{align}
for all $g \in \mathfrak{g}$ and $h, h^{\prime} \in \mathfrak{h}$.
Since $\undert$ and $\underalpha$ are homomorphims, we get
\begin{align}
\undert([h, h^{\prime}])&=[\undert(h), \undert(h^{\prime})], \\
\underalpha([g, g^{\prime}])&=[\underalpha(g), \underalpha(g^{\prime})],
\label{dcrossedmodule0}
\\
\underalpha(g)([h, h^{\prime}])&= 
[\underalpha(g)h, h^{\prime}] + [h, \underalpha(g)h^{\prime}],
\label{differentialcrossedmodulehomo}
\end{align}
for all $g, g^{\prime} \in \mathfrak{g}$ and $h, h^{\prime} \in \mathfrak{h}$.
A differential crossed module is equivalent to a strict Lie 2-algebra.

Let us denote the bases of the Lie algebras by $g_a \in \mathfrak{g}$ and
$h_A \in \mathfrak{h}$. Their Lie brackets are 
\begin{align}
[g_{a},g_{b}]=&-f^{c}_{ab}g_{c},\
[h_{A},h_{B}]=\tilde{f}^{C}_{AB}h_{C},
\end{align}
with structure constants $f^{c}_{ab}$ and $\tilde{f}^{C}_{AB}$, respectively.
The maps $\undert$ and $\underalpha$ can be expressed as
\begin{align}
\undert(h_{A})&=t^{a}_{A}g_{a}, \\
\underalpha(g_{a})h_{A}&= \alpha_{aA}^{B}h_{B},
\end{align}
with coefficients $t^{a}_{A}$ and $\alpha_{aA}^{B}$, respectively. The structure constants $f^{c}_{ab}$, $\tilde{f}^{C}_{AB}$,
$t^{a}_{A}$ and $\alpha_{aA}^{B}$ satisfy 
the following relations,
\begin{align}
f^{d}_{e[a}f^{e}_{bc]} &=0, 
\label{dcrossedmodule1}\\
\tilde{f}^{D}_{E[A}\tilde{f}^{E}_{BC]} &=0, \label{dcrossedmodule2} \\
[g_{b},\undert(h_{A})]-\undert(\underalpha(g_{b})h_{A}) 
&= t^{c}_{A}f^{a}_{cb}g_{a}-t^{a}_{B}\alpha^{B}_{bA}g_{a}
= 0, \label{dcrossedmodule3} \\
\underalpha(\undert(h_{A}))h_{B}-[h_{A},h_{B}] &=\alpha^{C}_{aB}t^{a}_{A}h_{C}-\tilde{f}^{C}_{AB}h_{C}
=0. \label{dcrossedmodule4}
\end{align}
From \eqref{dcrossedmodule4}, we obtain
\begin{align}
\tilde{f}^{C}_{AB}=t^{a}_{A}\alpha^{C}_{aB}.
\label{dcrossedmodule5}
\end{align}
Therefore, $\tilde{f}^{C}_{AB}$ is expressed by
$t^{a}_{A}$ and $\alpha^{C}_{aB}$,
and the conditions of the differential crossed module can be 
written by $f^c_{ab}$, 
$t^{a}_{A}$ and $\alpha^{C}_{aB}$ only.
Since $\tilde{f}^{C}_{AB}$ is antisymmetric,
we obtain
\begin{align}
t^{a}_{B}\alpha^{C}_{aA}+t^{a}_{A}\alpha^{C}_{aB}&=0.
\label{dcrossedmodule6}
\end{align}
Using \eqref{dcrossedmodule0}, we find
\begin{align}
\alpha^{B}_{cA}f^{c}_{ab}+\alpha^{B}_{aC}\alpha^{C}_{bA}
-\alpha^{B}_{bC}\alpha^{C}_{aA}&=0.
\label{dcrossedmodule7}
\end{align}
Finally, we can summarize all conditions as follows,
\begin{align}
f^{d}_{e[a}f^{e}_{bc]} &=0,\\
t^{c}_{A}f^{a}_{cb}-t^{a}_{B}\alpha^{B}_{bA} &=0,\\
\alpha^{B}_{cA}f^{c}_{ab}+\alpha^{B}_{aC}\alpha^{C}_{bA}
-\alpha^{B}_{bC}\alpha^{C}_{aA}&=0,\\
t^{a}_{B}\alpha^{C}_{aA}+t^{a}_{A}\alpha^{C}_{aB}&=0.
\end{align}
These equations reproduce a differential crossed module.
\medskip

A \emph{semistrict Lie 2-algebra}, which is a generalization
of a differential crossed module, is a pair of vector spaces
$\mathfrak{g}$ and $\mathfrak{h}$ with the following operations:
An antisymmetric 2-bracket 
$[-,-]: \mathfrak{g} \times \mathfrak{g} \rightarrow \mathfrak{g}$,
a totally antisymmetric 3-bracket 
$[-,-,-]: \mathfrak{g} \times \mathfrak{g} 
\times \mathfrak{g} \rightarrow \mathfrak{h}$
and two maps
$\undert: \mathfrak{h} \rightarrow \mathfrak{g}$
and
$\underalpha(-): \mathfrak{g} \times \mathfrak{h} \rightarrow \mathfrak{h}$.
These operations satisfy
\begin{align}
& [g, \undert(h)] = \undert(\underalpha(g)h), \\
& \underalpha(\undert(h))h^{\prime} = - \underalpha(\undert(h^{\prime}))h, \\
& [g_1,[g_2,g_3]]+ [g_2,[g_3,g_1]]+ [g_3, [g_1, g_2]] = \undert([g_1, g_2, g_3]), \\
& \underalpha(g_1)\underalpha(g_2)h - \underalpha(g_2)\underalpha(g_1)h
- \underalpha([g_1, g_2]) h = [g_1, g_2, \undert(h)], \\
& \underalpha(g_1)[g_2, g_3, g_4] - \underalpha(g_2)[g_3, g_4, g_1]
+ \underalpha(g_3)[g_4, g_1, g_2] - \underalpha(g_4)[g_1, g_2, g_3]
\nonumber \\
&
- [g_1, g_2, [g_3, g_4]] - [g_1, g_3, [g_4, g_2]] - [g_1, g_4, [g_2, g_3]] 
\nonumber \\
&
+ [g_2, g_3, [g_4, g_1]] + [g_4, g_2, [g_3, g_1]] + [g_3, g_4, [g_2, g_1]] 
= 0,
 \label{semistrictlie2algebra}
\end{align}
for all $g_i \in \mathfrak{g}$ and $h_i \in \mathfrak{h}$.
If $[-,-,-] =0$, the semistrict Lie 2-algebra becomes a strict Lie 2-algebra.

If we choose bases $g_a \in \mathfrak{g}$ and
$h_A \in \mathfrak{h}$, then each operation can be expressed by
\begin{align}
[g_{a},g_{b}]=&-f^{c}_{ab}g_{c},\\
\undert(h_{A})&=t^{a}_{A}g_{a}, \\
\underalpha(g_{a})h_{A}&= \alpha_{aA}^{B}h_{B}. \\
[g_a,g_b,g_c] &=T^{A}_{abc} h_{A}.
\end{align}
The structure constants $f^{c}_{ab}$, $t^{a}_{A}$, $\alpha_{aA}^{B}$ and $T^{A}_{abc}$ satisfy the following relations,
\begin{align}
\frac{1}{2}f^{d}_{e[a}f^{e}_{bc]}&-\frac{1}{3!}t^{d}_{A}T^{A}_{abc}=0,\\
t^{c}_{A}f^{a}_{cb}&-t^{a}_{B}\alpha^{B}_{bA}=0,\\
\frac{1}{2}\alpha^{B}_{cA}f^{c}_{ab}&+\alpha^{B}_{[a|C|}\alpha^{C}_{b]A}+\frac{1}{2}t^{c}_{A}T^{B}_{cab}=0,\\
\frac{3}{2}f^{e}_{[ab}T^{A}_{cd]e}&+\alpha^{A}_{[a|B|}T^{B}_{bcd]}=0,\\
\alpha^{C}_{a(A}t^{a}_{B)}&=0.
\end{align}

\section{Special solutions in 5 dimensions}\label{sec:CaseIII}

In this subsection, we derive special solutions of the master equation $\{\Theta^{(1)} + \Theta^{(2)}, \Theta^{(1)} + \Theta^{(2)}\}$ with $T^A_{abc}=0$. 
For this, we assume that $G^{ab}\equiv t^{a}_{A}s^{bA}$ is not invertible, in general.
Furthermore, we assume that there exists an invertible metric $g_{ab}$ on $W$. 
We define $s_{a}^{A} \equiv g_{ab}s^{bA}$ and introduce the matrix ${\cal P}_{b}^{a}=t^{b}_{A}s_{a}^{A}$. We assume, that
\begin{equation}
s_{a}^{A}t^{a}_{B}=\delta^{A}_{B},\label{eq:rj384wpguinvljkdf}
\end{equation}
so that ${\cal P}$ is a projection, ${\cal P}^{a}_{b}{\cal P}^{b}_{c}={\cal P}^{a}_{c}$.
Then, from (\ref{condition7}), we obtain
\begin{align}
n^{AB}_{a}=s^{B}_{b}s^{cA}f^{b}_{ca}+s^{B}_{b}s^{bC}\alpha^{A}_{aC}.
\end{align}
Using the crossed module relations, we can represent $\alpha^A_{aB}$ as
\begin{align}
\alpha^{A}_{aB}=s^{A}_{b}t_{B}^{c} f^{b}_{ca}.\label{eq:bnguorgtruioh}
\end{align}
This relation leads to
\begin{align}
n^{AB}_{a}=2s^{c(A}s^{B)}_{b}f^{b}_{ca}.\label{eq:wqyuyrui4yriuyriu}
\end{align}
Next, we show that the expression for $n^{AB}_{a}$ satisfies all other equations.
First, by multiplication of \eqref{eq:wqyuyrui4yriuyriu} by $s^{aC}$ we obtain 
\begin{align}
s^{aC}n^{AB}_{a}=s^{aC}(s^{cA}s^{B}_{b}f^{b}_{ca}+s^{cB}s^{A}_{d}f^{d}_{ca})=2s^{a[C}s^{c|A]}s^{B}_{b}f^{b}_{ca}+2s^{a[C}s^{c|B]}s^{A}_{d}f^{d}_{ca}.
\end{align}
Therefore, we find $s^{a(A}n_a^{BC)}=0$.
Second, from (\ref{condition9}) we derive
\begin{align}
\frac{1}{2}t^{a}_{C}n^{AB}_{a}=t^{a}_{C}s^{c(A}s^{B)}_{b}f^{b}_{ca}=-s^{c(A}s^{B)}_{b}t^{a}_{C}f^{b}_{ac}=-s^{c(A}\alpha^{B)}_{cC},
\end{align}
where we used (\ref{eq:bnguorgtruioh}). Therefore, equation (\ref{condition9}) holds.
The equation (\ref{condition8}) gives an additional condition. Using (\ref{eq:wqyuyrui4yriuyriu}), we get
\begin{equation}
\frac{1}{4}n^{AB}_{c}f^{c}_{ab} =s^{d(A}s^{B)}_{e}f^{e}_{c[a}f^{c}_{b]d}.
\end{equation}
Similarly, we derive
\begin{equation}
\alpha^{A}_{aC}n^{BC}_{b} =s^{A}_{e}s^{cB}f^{e}_{fa}{\cal P}^{f}_{d}f^{d}_{cb}+s^{A}_{e}s^{B}_{d}f^{e}_{fa}{\cal P}^{fc}f^{d}_{cb}.
\end{equation}
Combining both results leads to
\begin{align}
\alpha^{(A}_{[aC}n^{B)C}_{b]}=-s^{(A}_{e}s^{cB)}{\cal P}^{f}_{d}f^{e}_{f[a}f^{d}_{b]c}=-s^{d(A}s^{B)}_{e}{\cal P}^{f}_{c}f^{e}_{f[a}f^{c}_{b]d}.
\end{align}
Finally, we obtain
\begin{align}
\frac{1}{4}n^{AB}_{c}f^{c}_{ab}+\alpha^{(A}_{[aC}n^{B)C}_{b]}
&=s^{d(A}s^{B)}_{e}(\delta^{f}_{c}-{\cal P}^{f}_{c})f^{e}_{f[a}f^{c}_{b]d},\nonumber\\
&=g^{dg}s^{(A}_{g}\{s^{B)}_{e}(\delta^{f}_{c}-{\cal P}^{f}_{c})f^{e}_{f[a}f^{c}_{b]d}\}
\end{align}
Therefore, we get the following additional condition,
\begin{align}
g^{dg}s^{(A}_{g}\{s^{B)}_{e}(\delta^{f}_{c}-{\cal P}^{f}_{c})f^{e}_{f[a}f^{c}_{b]d}\}=0,\label{eq:bnsjkdxur8349wsgibnjxd}
\end{align}
which is satisfied in the example we used for off-shell covariantization in 5 dimensions.

%% file: app02.tex
\end{comment}

\section{Master equation on ${\cal M}_4$}\label{Appendixn4}
In this section, we show the general results of the calculation of the classical master equation on $T^*[4]{\cal N}$, where ${\cal N} = E[1] \oplus E^{\prime}[2]$. $E$ and $E'$ are  vector bundles over a smooth manifold $M$. Note that this gives a symplectic Lie 4-algebroid, since we are allowing fibrations over a manifold. In the main text, we worked with a sympletic Lie 4-algebra. 
The local coordinates of $T^{*}[4]{\cal N}$ are
\begin{align}
(x^i, q^{a},Q^{A}), \quad (\xi_{i},p_{a},P_{A})
\end{align}
of degree $(0, 1, 2)$ and $(4, 3, 2)$.
The canonical Poisson bracket is defined by 
\begin{align}
\{f,g\}
=\frac{\partial f}{\partial x^{i}}\frac{\partial g}{\partial \xi_{i}}-\frac{\partial f}{\partial \xi_{i}}\frac{\partial g}{\partial x^{i}}+\frac{\partial f }{\partial q^{a}}\frac{\partial g}{\partial p_{a}}+\frac{\partial f}{\partial p_{a}}\frac{\partial g}{\partial q^{a}}+\frac{\partial f}{\partial Q^{A}}\frac{\partial g}{\partial P_{A}}-\frac{\partial f}{\partial P_{A}}\frac{\partial g}{\partial Q^{A}}.
\end{align}
The Hamiltonian function is given by
\begin{equation}
\Theta=\Theta^{(0)}+\Theta^{(1)}+\Theta^{(2)},
\end{equation}
where
\begin{align}
\Theta^{(1)}&=\tau^{i}_{a}q^{a}\xi_{i}+\frac{1}{2}f^{c}_{ab}q^{a}q^{b}p_{c}+t^{a}_{A}Q^{A}p_{a}+\alpha^{B}_{aA}q^{a}Q^{A}P_{B}+\frac{1}{3!}T^{A}_{abc}q^{a}q^{b}q^{c}P_{A},\label{eq:Vongolebiancod}\\
\Theta^{(0)}&=\frac{1}{5!}m_{abcde}q^{a}q^{b}q^{c}q^{d}q^{e}+\frac{1}{3!}m_{abcA}q^{a}q^{b}q^{c}Q^{A}+\frac{1}{2}m_{aAB}q^{a}Q^{A}Q^{B},\label{eq:pescatorad}\\
\Theta^{(2)}&=s^{aA}p_{a}P_{A}+\frac{1}{2}n_{a}^{AB}q^{a}P_{A}P_{B}.\label{eq:Genovesed}
\end{align}
The classical master equation induces the following equations,
\begin{align}
-\tau^{i}_{[a}\partial_{i}\tau^{j}_{b]}&+\frac{1}{2}\tau^{j}_{c}f^{c}_{ab}=0,\\
-\frac{1}{2}\tau^{i}_{[a}\partial_{i}f^{d}_{bc]}&+\frac{1}{2}f^{d}_{e[a}f^{e}_{bc]}-\frac{1}{3!}t^{d}_{A}T^{A}_{abc}+\frac{1}{3!}s^{dA}m_{abcA}=0,\\
-\tau^{i}_{b}\partial_{i}t^{a}_{A}&+t^{c}_{A}f^{a}_{cb}-t^{a}_{B}\alpha^{B}_{bA}+s^{aB}m_{bBA}=0,\\
-\tau^{i}_{[a}\partial_{i}\alpha^{B}_{b]A}&+\frac{1}{2}\alpha^{B}_{cA}f^{c}_{ab}+\alpha^{B}_{[a|C|}\alpha^{C}_{b]A}+\frac{1}{2}t^{c}_{A}T^{B}_{cab}+\frac{1}{2}s^{cB}m_{cabA}-n_{[a}^{CB}m_{b]CA}=0,\\
\tau^{i}_{[a}\partial_{i}T^{A}_{bcd]}&+\frac{3}{2}f^{e}_{[ab}T^{A}_{cd]e}+\alpha^{A}_{[a|B|}T^{B}_{bcd]}+\frac{1}{4}s^{eA}m_{eabcd}+n^{AB}_{[a}m_{bcd]B}=0,\\
\tau^{i}_{a}t^{a}_{A}&=0,\\
\alpha^{C}_{a(A}t^{a}_{B)}&+\frac{1}{2}s^{aC}m_{aAB}=0,
\end{align}
\begin{align}
s^{a(A}n^{BC)}_{a}=0,
\end{align}
\begin{align}
-\tau^{i}_{a}\partial_{i}s^{bA}&+s^{cA}f^{b}_{ca}+\alpha^{A}_{aB}s^{bB}-t^{b}_{B}n^{AB}_{a}=0,\\
-\frac{1}{2}\tau^{i}_{[a}\partial_{i}n_{b]}^{AB}&+\frac{1}{2}s^{c(A}T^{B)}_{abc}+\frac{1}{4}n^{AB}_{c}f^{c}_{ab}+\alpha^{(A}_{[a|C|}n^{B)C}_{b]}=0,\\
s^{a(A}\alpha^{B)}_{aC}&+\frac{1}{2}t^{a}_{C}n^{AB}_{a}=0,\\
\tau^{i}_{a}s^{aA}&=0,\\
t^{[a}_{A}s^{b]A}&=0,
\end{align}
\begin{align}
\tau^{i}_{[f}\partial_{i}m_{abcde]}&+\frac{5}{2}f^{g}_{[ef}m_{abcd]g}+\frac{10}{3}m_{[abc|A|}T^{A}_{def]}=0,\\
\tau^{i}_{[d}\partial_{i}m_{abc]A}&+\frac{1}{4}t^{e}_{A}m_{eabcd}+\frac{3}{2}f^{e}_{[cd}m_{ab]eA}+T^{B}_{[bcd}m_{a]BA}+m_{[abc|B|}\alpha^{B}_{d]A}=0,\\
-\tau^{i}_{[a}\partial_{i}m_{b]AB}&+\frac{1}{4}f^{c}_{ab}m_{cAB}+\frac{1}{2}t^{c}_{(A}m_{|cab|B)}+\frac{1}{2}(m_{aC(A}\alpha^{C}_{|b|B)}-m_{bC(A}\alpha^{C}_{|a|B)})=0,\\
m_{a(AB}t^{a}_{C)}&=0.
\end{align}


%% file: main.bbl
\begin{thebibliography}{10}

\bibitem{Ho:2012nt}
P.-M. Ho and Y.~Matsuo,
\newblock JHEP {\bf 09}, 075 (2012), 1206.5643.

\bibitem{Witten:1995zh}
E.~Witten,
\newblock {Some comments on string dynamics},
\newblock in {\em {Future perspectives in string theory. Proceedings,
  Conference, Strings'95, Los Angeles, USA, March 13-18, 1995}}, 1995,
  hep-th/9507121.

\bibitem{Witten:2009at}
E.~Witten,
\newblock (2009), 0905.2720.

\bibitem{Lambert:2010wm}
N.~Lambert and C.~Papageorgakis,
\newblock JHEP {\bf 08}, 083 (2010), 1007.2982.

\bibitem{Bonetti:2012st}
F.~Bonetti, T.~W. Grimm, and S.~Hohenegger,
\newblock JHEP {\bf 05}, 129 (2013), 1209.3017.

\bibitem{Chu:2012um}
C.-S. Chu and S.-L. Ko,
\newblock JHEP {\bf 05}, 028 (2012), 1203.4224.

\bibitem{Ho:2014eoa}
P.-M. Ho and Y.~Matsuo,
\newblock JHEP {\bf 12}, 154 (2014), 1409.4060.

\bibitem{Ritter:2015zur}
P.~Ritter, C.~Saemann, and L.~Schmidt,
\newblock (2015), 1512.07554.

\bibitem{Ho:2011ni}
P.-M. Ho, K.-W. Huang, and Y.~Matsuo,
\newblock JHEP {\bf 07}, 021 (2011), 1104.4040.

\bibitem{Perry:1996mk}
M.~Perry and J.~H. Schwarz,
\newblock Nucl. Phys. {\bf B489}, 47 (1997), hep-th/9611065.

\bibitem{Pasti:1996vs}
P.~Pasti, D.~P. Sorokin, and M.~Tonin,
\newblock Phys. Rev. {\bf D55}, 6292 (1997), hep-th/9611100.

\bibitem{Henneaux:1997ha}
M.~Henneaux and B.~Knaepen,
\newblock Phys. Rev. {\bf D56}, 6076 (1997), hep-th/9706119.

\bibitem{Bekaert:2000qx}
X.~Bekaert, M.~Henneaux, and A.~Sevrin,
\newblock Commun. Math. Phys. {\bf 224}, 683 (2001), hep-th/0004049.

\bibitem{Baez:2010ya}
J.~C. Baez and J.~Huerta,
\newblock Gen. Rel. Grav. {\bf 43}, 2335 (2011), 1003.4485.

\bibitem{Baez:2003fs}
J.~C. Baez and A.~S. Crans,
\newblock Theor. Appl. Categor. {\bf 12}, 492 (2004), math/0307263.

\bibitem{Palmer:2014jma}
S.~Palmer,
\newblock {\em {Higher Gauge Theory and M-Theory}},
\newblock PhD thesis, Heriot-Watt U., 1407.0298.

\bibitem{vanNieuwenhuizen:1982zf}
P.~van Nieuwenhuizen,
\newblock {FREE GRADED DIFFERENTIAL SUPERALGEBRAS},
\newblock in {\em {11th International Colloquium on Group Theoretical Methods
  in Physics (GROUP 11) Istanbul, Turkey, August 23-28, 1982}}.

\bibitem{D'Auria:1982pm}
R.~D'Auria, P.~Fre, P.~K. Townsend, and P.~van Nieuwenhuizen,
\newblock Annals Phys. {\bf 155}, 423 (1984).

\bibitem{Bojowald:2004wu}
M.~Bojowald, A.~Kotov, and T.~Strobl,
\newblock J. Geom. Phys. {\bf 54}, 400 (2005), math/0406445.

\bibitem{Lavau:2014iva}
S.~Lavau, H.~Samtleben, and T.~Strobl,
\newblock J. Geom. Phys. {\bf 86}, 497 (2014), 1403.7114.

\bibitem{Gruetzmann:2014ica}
M.~Gr{\"u}tzmann and T.~Strobl,
\newblock Int. J. Geom. Meth. Mod. Phys. {\bf 12}, 1550009 (2014), 1407.6759.

\bibitem{Schwarz:1992nx}
A.~S. Schwarz,
\newblock Commun. Math. Phys. {\bf 155}, 249 (1993), hep-th/9205088.

\bibitem{Schwarz:1992gs}
A.~S. Schwarz,
\newblock Commun. Math. Phys. {\bf 158}, 373 (1993), hep-th/9210115.

\bibitem{Castellani}
R.~D. L.~Castellani and P.~Fr\'e,
\newblock {\em {Supergravity and superstrings vol. 1,2,3}} (World Scientific,
  1991).

\bibitem{Fiorenza:2010mh}
D.~Fiorenza, U.~Schreiber, and J.~Stasheff,
\newblock Adv. Theor. Math. Phys. {\bf 16}, 149 (2012), 1011.4735.

\bibitem{Bessho:2015tkk}
T.~Bessho, M.~A. Heller, N.~Ikeda, and S.~Watamura,
\newblock (2015), 1511.03425.

\bibitem{Alexandrov:1995kv}
M.~Alexandrov, M.~Kontsevich, A.~Schwartz, and O.~Zaboronsky,
\newblock Int. J. Mod. Phys. {\bf A12}, 1405 (1997), hep-th/9502010.

\bibitem{Ikeda:2012pv}
N.~Ikeda,
\newblock (2012), 1204.3714.

\bibitem{Liu-Weinstein-Xu}
Z.-J. Liu, A.~Weinstein, and P.~Xu,
\newblock Journal of Differential Geometry {\bf 45}, 547 (1997).

\bibitem{Roytenberg:1999}
D.~Roytenberg,
\newblock {\em Courant algebroids, derived brackets and even symplectic
  supermanifolds} (, 1999), math/9910078,
\newblock Thesis (Ph.D.)--University of California, Berkeley.

\bibitem{Baez:2002jn}
J.~C. Baez,
\newblock (2002), hep-th/0206130.

\end{thebibliography}
